\documentclass{article}

\usepackage[utf8]{inputenc}
\usepackage{amssymb,amsthm}
 \usepackage[english]{babel}
\usepackage{subcaption}
\usepackage[noblocks]{authblk}

\usepackage[letterpaper,top=2cm,bottom=2cm,left=3cm,right=3cm,marginparwidth=1.75cm]{geometry}
\usepackage{amsmath, amssymb, amsfonts, amsthm}
\usepackage{graphicx}
\usepackage[colorlinks=true, allcolors=blue]{hyperref}

\newcommand{\Rr}{\mathbb{R}}

\newcommand{\Ss}{\mathbb{S}}
\newcommand{\Tt}{\mathbb{T}}

\newcommand{\SU}{\mathfrak{su}}

\DeclareMathOperator{\Tr}{Tr}

\newtheorem{thm}{Theorem}

\newtheorem{cor}[thm]{Corollary}
\newtheorem{lemma}[thm]{Lemma}

\newtheorem{rmk}[thm]{Remark}

\title{The early stage of the motion along the gradient of a concentrated vortex structure}
\author[*]{Franco Flandoli} \author[**]{Matteo Palmieri} \author[*]{Milo Viviani}
\affil[*]{Scuola Normale Superiore, Pisa \linebreak \textit{franco.flandoli@sns.it} and \textit{milo.viviani@sns.it}}
\affil[**]{Max Planck Institute for Mathematics in the Sciences, Leipzig \linebreak \textit{matteo.palmieri@mis.mpg.de}}

\date{}

\begin{document}
\maketitle

\begin{abstract}
We give a rigorous mathematical result, supported by numerical simulations, of the aggregation of a concentrated vortex blob with an underlying non-constant vorticity field: the blob moves in the
direction of the gradient of the field. It is a unique example of a Lagrangian
explanation of aggregation of vortex structures of the same sign in 2D
inviscid fluids. The result is also extended to almost vertical vortex filaments in a (possibly thin) three-dimensional domain.
\end{abstract}

\section{Introduction}

Vortex structures emerge from many instabilities mechanisms of fluid and
plasma dynamics, like Rayleigh-Taylor, Richtmyer--Meshkov, and
Kelvin--Helmholtz instabilities \cite{Zhou2017}, \cite{Zhou2024},
\cite{Zhou2025}. In the case of 2D (or nearly two-dimensional) fluids, their
interaction and in particular their merging properties play a fundamental role
in many processes, like in the turbulent inverse cascade \cite{Boffetta}, in
the emergence and dynamics of large scale structures in the atmosphere of
certain planets \cite{Rhines}, \cite{Vasavada} and in the poloidal section of
confined fusion plasma \cite{Diamond}, \cite{Hasegawa}. Here and below we
shall generically name it vorticity, meaning true vorticity for classical
fluids, but potential vorticity for geophysical fluids, \cite{Vallis} and a
particle\ density or the electrostatic potential - depending on the
approximate model - for a plasma \cite{Briggs}, \cite{Horton}.

With a large degree of simplification, we may divide the mutual interaction of
vortex structures and their interaction with background fields into two main classes.

a) The first class contains the interaction of vortex structures approximately
of the same size (with small variants of different size but with the same
order of magnitude \cite{Ozugurlu}). The present paper is not devoted to this
case, but we mention it for comparison and side remarks. When the two vortex
structures have the same sign and are sufficiently close to each other, they
merge into a larger structure (hence the energy previously incorporated in the
smaller ones moves to the large final one, as the inverse energy cascade
predicts) and produce very small scale structures like filaments (hence part
of the vorticity, previously incorporated in the two initial structures, will
be found in the very thin filaments, in agreement with the direct enstrophy
cascade \cite{Krajchnan}). This process has been widely investigated, in
particular for the potential vorticity field in geostrophic dynamics and
turbulence\textit{\ }\cite{Hardenberg}, \cite{Nielsen}, \cite{Reinaud}, and
has been experimentally observed and investigated in the case of the oval BA (Red Spot
Jr.) of Jupiter \cite{Sanchez}, \cite{Marcus2000}, \cite{Marcus2004},
\cite{Wong}; but it has been also studied for simplified plasma models, see for instance \cite{Ogawa} and references therein. 

b)\ The second class contains the case of interaction of vortex structures of
very different sizes, so that one looks almost pointwise compared to the
other. Similarly, it contains the case of the interaction between an almost
pointwise structure and a background shear flow. It is called the vortex-wave
problem in the limit of a pointwise vortex \cite{MarcPulv}; the small vortex
structures are also often called clumps and holes \cite{Dupree}. In this
category, we may also insert the beta-drift problem, namely the motion of a
concentrated vortex structure in the presence of Coriolis force, relevant for
instance to the motion of cyclones in the atmosphere \cite{Carnevale},
\cite{Chan}, \cite{Li}. The interaction of very concentrated vorticity
structures with much larger vortex structures or a vorticity background may be
important to understand the properties (e.g. energy maintenance, growth and
shaping) of large scale structures (like zonal flows or streamers) in planets
and plasma \cite{Vasavada}, \cite{Diamond}, \cite{Garbet}.

Case (a) is documented by several numerical studies, but it is less clear from
the theoretical point of view based on the equations of motion, with some
information extracted by statistical mechanics arguments related, for instance
, to the concept of entropy \cite{Dritschel}, \cite{Ozugurlu}, or Onsager theory
\cite{Flandoli}. As already said above, we have mentioned case (a) for
comparison, but it will not be the main object of this work.

Case (b) is more amenable to an analytical theory, because of the presence of
a small parameter (the ratio between the scales of the two structures) and because
in some regimes or small time one may assume a moderate modification of the
background. We focus here on this case. More precisely, we focus on a specific
property of the interaction between a concentrated structure and a background:
the \textit{motion of the structure along the gradient of the background
vorticity.} Beyond being of interest in itself, this particular property is in
a loose sense a property of \textit{aggregation} of vorticity structures
having the same sign, thus with some analogy with the coalescence of similar
size vortices mentioned in point (a)\ above. Moreover, there is some loose
analogy also in some detail of the process, like the formation of a the strong
shear layer between the structures during the merging process (a)
\cite{Nielsen}, \cite{Ogawa} or the encapsulation process (b) \cite{Amoretti},
\cite{Lansky}.

We do not discuss in detail the rich literature focusing on other questions
for (b) different from the question of the motion along the gradient. Among
others, let us only mention the analysis of the oscillations of the
concentrated vortex when it is close to the boundary of a large one and could
lead to coalescence of the two \cite{Lansky} (containing an interesting
hamiltonian analysis of the model); or the growth of holes and their dynamics
\cite{Dupree} which contains many ideas and methods used later on by other authors.

We specifically recall, on the contrary, the experimental and numerical
investigation of the motion along the gradient performed by Amoretti et al.
\cite{Amoretti} which highlights the complexity of the interaction layers
between concentrated vortex and the interior of the large one. And the deep
analytical analysis of the motion along the gradient by Schecter and Dubin
\cite{Schecter}, who found quantitative formulae for the motion, with good
numerical fit in some case, and explain well also the heuristics behind the
methodology, based on the idea of a strong local mixing of the background near
the concentrated vortex, which produces a motion of the vortex by conservation
of momentum.

In spite of the previously mentioned progress, which provided sure evidence
that a motion along the gradient exists and understood some of its features, a
rigorous mathematical proof, without approximations, does not exist. For
instance, the analytical computation of \cite{Schecter} is based on a sequence
of two approximations:\ first a formula for the approximate deformation of the
background field produced by the very strong action of the concentrated vortex
structure; then an approximate formula for the displacement caused by the
deformed field on the structure. These approximations are very realistic and
implementing them in sequence instead of in parallel (in spite of the fact
that the two actions happen simultaneously) looks an acceptable approximation
in view of the difficulty of the problem. However, we see two criticalities.
The first one is the lack of rigorous proof. Maybe more important, the great
difficulty to extend the argument to problems which are similar but more
complex, where other approximations hold simultaneously (each of them should
then be under precise control for comparison, and it is not the case until
now). We have in mind in particular the \textit{quasi-two-dimensional case},
of paramount importance see \cite{Alexakis} and references therein. If the
same approximation arguments of previous works are applied to quasi-2D
structures, the link between the mathematical approximations, their 3D effects
(potentially very strong) and the small parameters characterizing the quasi-2D
set-up should be very clear.

Our primary aim in this paper is adding some information on the
\textit{initial stage} of the motion along the gradient. The initial stage is
cleaner, since the background flow at time zero is assumed to be a shear flow.
Later stages may be very complex, namely the background may suffer a strong
modification. In certain experiments and real world problems the shear-flow
structure of the background could be reinforced by some mechanism, the shear
flow structure may be restored to some degree by other forces acting on the
fluid (for instance a pressure difference maintained between two positions),
and these factors could contribute to prolong the time where we observe a
clean motion along the gradient. But in the numerical simulations of the basic
model equations (the Euler equations without forces), the clump (or hole)
typically has a strong influence on the background which alters too much its
structure and makes quite unpredictable and complex the dynamics after short
time;\ also the concentrated vortex structure undergoes transformations after
short time \cite{Mariotti}. Therefore we put ourselves in the idealized
situation of a clump in a strict shear flow at time zero, and aim to
understand the first dynamical phase, before it reaches an uncontrolled
complexity. With this restriction to the first phase of the dynamics, our aim
is to develop a completely rigorous analysis, based on the full nonlinear
equations of motion, free of any assumption and approximation.

Among the advantages of an approach free from uncontrolled approximations,
there is potentiality to be extended to more complex cases. Here we shall
investigate a quasi-two-dimensional framework, namely a vortex filament in a
3D shear flow, under assumptions of small deviation from a 2D case. The
equations for vortex filaments have a long story, are notoriously difficult
and present potential singularities. The question investigated here, of the
motion along the gradient of a background shear flow, is new but other aspects
of the motion of filaments, like well posedness and singularities, has been
studied for long time, see for instance \cite{Darios}, \cite{LeviCivita},
\cite{Saffman}, \cite{Keener}, \cite{Ricca}, \cite{Fukumoto}, \cite{Majda}%
,\ \cite{Benedetto}, \cite{DelaHoz}, \cite{Jerrard}, \cite{Banica1}%
,\ \cite{Banica2},\ \cite{Fontelos} and other references in the technical
sections below.

Concerning mathematically rigorous approaches to vortex dynamics, but for
questions different from the one treated here, we are indebt to important
contributions to non-pointwise vortex structures \cite{Butta}, \cite{Majda},
\cite{Marchioro} and to the dynamics of point vortices and its statistical
mechanics properties, among others \cite{Marchioro}, \cite{Caglioti},
\cite{Chavanis}, \cite{Eyink}, \cite{Eyink2}, \cite{Lions}, \cite{Montgomery},
\cite{Onsager}. These results are not directly related to our work but they
are generically in the background of our investigation. Other more directly
related references will be quoted below in the technical sections.

The numerical simulations performed in the literature on these problems are
various, with some preference for the particle-in-cell method \cite{Amoretti}.
We shall explore the performances of a variant of the Zeitlin model, on the
rotating 2D sphere, that we call the vortex-wave Zeitlin model aiming to
confirm the theoretical predictions.

In our problem the small intense vortex is not generated by the shear flow instability; the shear flow is still unperturbed at time $t = 0$, when the merging process starts. Clearly, the intrusion of the intense vortex in the shear flow will
produce a strong perturbation of the latter, but we do not aim to give rigorous results on the evolution of the background perturbation, only on the initial-stage motion of the concentrated vortex in the direction of the initial gradient of the
background vorticity.

Vortices in shear flow have also been considered from another viewpoint, see
\cite{MooreSaffman}. In that work, the authors found stable vortices stretched by the flow. But the basic assumption in that case is that the shear is constant, namely the vorticity is constant. Our result of motion along the
vorticity gradient requires a non-zero vorticity gradient.

\section{Local dynamics of the blob-wave system }\label{blob}

In this section, for the sake of clarity of the calculations, we take as reference domain the flat torus $\mathbb{T}^2$. The Euler equation is the following nonlinear transport equation for the vorticity $\omega^t:\mathbb{T}^2\to\mathbb{R}$ of fluid, advected by the velocity field $\mathbf{u}^t=(u_1^t,u_2^t):\mathbb{T}^2\to\mathbb{R}$ it generates
\begin{align}\label{e8}
\partial_t\omega+\mathbf{u}\cdot\nabla\omega=0\qquad\mathbf{u}=\mathbf{K}\ast\omega,
\end{align}
where $\mathbf{K}(\mathbf{x},\mathbf{y}):=\nabla^\perp\Delta_{\mathbb{T}}^{-1}\delta_{\mathbf{y}}(\mathbf{x})$ denotes the Biot-Savart kernel on the torus. The initial value problem is well-posed for every bounded initial datum
\begin{align*}
\omega^0(\mathbf{x})\in L^\infty(\mathbb{T}^2)
\end{align*}
and produces a unique solution $\omega\in L^\infty([0,\infty)\times\mathbb{T}^2)$ (see \cite{Yudovich}). We will work in this setting.

The aim of this section is to rigorously prove some results concerning the dynamics of a concentrated blob of vorticity with respect to a shear flow with a non-zero vorticity gradient. Let $\bar{\omega}^0(\mathbf{x})$ be a steady smooth solution $\bar{\mathbf{u}}^0$ be the associated velocity field. This is equivalent (cf.~\eqref{e8}) to requiring that
\begin{align}\label{e24}
\mbox{the vectors $\bar{\mathbf{u}}^0(\mathbf{x})$ and $\nabla\bar{\omega}^0(\mathbf{x})$ are orthogonal}\qquad\mbox{for every $\mathbf{x}\in\mathbb{T}^2$.}
\end{align}

A passive particle immersed in the fluid moves according to the differential equation $\dot{\mathbf{x}} = \bar{\mathbf{u}}^0(\mathbf{x})$. A point vortex with non-zero vorticity modifies the underlying vorticity field $\bar{\omega}^0$, so that its motion deviates from that of a passive particle. 
We want to describe this deviation, namely the motion along the direction perpendicular to $\bar{\mathbf{u}}^0(\mathbf{x})$, so parallel to $\nabla\bar{\omega}^0(\mathbf{x})$ (whenever this vector is non-zero).

To make our calculation rigorous we will not use the exact vortex-wave model (see the later section~\ref{s:vw}), but instead a \emph{blob-wave} system of equations, where we substitute the delta Dirac with a smooth, compactly supported function with a diameter of the support of order $\varepsilon \ll 1$. To be more precise, let $\mathbf{x}^0\in\Tt^2$ (the center of the blob), $\Gamma\in\Rr$ (the circulation the blob) and $p(\mathbf{x})\in C^\infty_c(\Rr^2)$ be a smooth function such that\footnote{by $dx:=dx_1dx_2$ we mean the volume element of the torus.}
\begin{align*}
p\geq0,\qquad\int p(\mathbf{x}) d x=1,\qquad\int\mathbf{x}~p(\mathbf{x}) d x=0,\qquad\text{and}\qquad \text{supp}~p \subset B(0,\frac{1}{2}).
\end{align*}
For $\varepsilon>0$, define
\begin{align}\label{e30}
\rho^{0,\varepsilon}(\mathbf{x})=\Gamma p^\varepsilon(\mathbf{x}):=\frac{\Gamma}{\varepsilon^2} p\left(\frac{\mathbf{x}-\mathbf{x}^0}{\varepsilon}\right).
\end{align}
which will be the profile of a vortex around $\mathbf{x}^0$ with total circulation $\Gamma$. The blob-wave dynamics is the solution of the following problem
\begin{align}\label{e25}
\begin{cases}
\partial_t\bar{\omega}+\mathbf{u}\cdot\nabla\bar{\omega}=0\qquad \bar{\omega}^{t=0}(\mathbf{x})=\bar{\omega}^0(\mathbf{x}),\\
\partial_t\rho+\mathbf{u}\cdot\nabla\rho=0\qquad\rho^{t=0}(\mathbf{x})= \rho^{0,\varepsilon}(\mathbf{x}) \\
\mathbf{u}^t(\mathbf{x})=\displaystyle\int_{\Tt^d}\mathbf{K}(\mathbf{x},\mathbf{y})(\bar{\omega}^t(\mathbf{y})+\rho^t(\mathbf{y}))  d y
\end{cases}  
\end{align}
where $\bar{\omega}^t(\mathbf{x})$ and $\rho^t(\mathbf{x})$ describe respectively the evolution of the large-scale vorticity field $\bar{\omega}$ and of the vortex $\rho$. Summing the first two equations, we obtain that the total vorticity $\omega = \bar{\omega} + \rho$ solves the Euler equations \eqref{e8} starting from the initial condition $\omega^0(\mathbf{x}) = \bar{\omega}^0(\mathbf{x}) + \rho^{0,\varepsilon}(\mathbf{x})$.

In \cite{MarcPulv} it is proven, in a slightly different setting, that the solutions of blob-wave dynamics converge (weakly as measures) to a solution of the vortex-wave system (Section~\ref{s:vw}), possibly allowing for a greater number of blobs, but with the additional assumption that the underlying vorticity field $\bar{\omega}^0$ is locally constant around the initial position of the blobs.

Following \cite{MarcPulv}, we introduce the barycenter of the blob
\begin{align*}
\Tt^2\ni \mathbf{G}(t)=\big(G_1(t),G_2(t)\big):=\int\mathbf{x}~\rho^t(\mathbf{x}) d x\Big/\int \rho^t(\mathbf{x}) d x=\frac{1}{\Gamma}\int\mathbf{x} ~\rho^t(\mathbf{x}) d x.
\end{align*}

\begin{thm}\label{thm1}
Assume that
\begin{align}\label{e28}
\mbox{$\bar{\omega}^0(\mathbf{x})$ depends only on $x_2$}\qquad\mbox{and}\qquad\Gamma>0,~\partial_2 \bar{\omega}^0(\mathbf{x}^0) > 0.
\end{align}
Then there exists a positive $\varepsilon_0 = \varepsilon_0(\bar{\omega}^0,\Gamma,p)$ and, for every $\varepsilon \in (0, \varepsilon_0]$, a time $T = T(\varepsilon) > 0$ such that $G_2(t)$ is stricly increasing in $[0, T]$.
More precisely,
\begin{align}\label{e26}
G_2'(0)=0\qquad\mbox{and}\qquad\Big|G_2''(0)-\frac{\Gamma\partial_2\bar\omega^0(\mathbf{x}^0)}{4\pi}\ln\frac{1}{\varepsilon}\Big|\lesssim_{\bar\omega^0,p}1.
\end{align}
\end{thm}

{\sc Proof.}
Introducing the velocity generated by the blob and the background vorticity respectively
\begin{align}\label{e29}
\mathbf{v}:=\mathbf{K}\ast\rho,\qquad\bar{\mathbf{u}}:=\mathbf{K}\ast\bar\omega.
\end{align}
we have
\begin{align*}
\partial_t \bar{\omega}^{t=0}(\mathbf{x})\overset{\eqref{e8}}{=}-\bar{\mathbf{u}}^0(\mathbf{x})\cdot\nabla\bar{\omega}^0(\mathbf{x}) - \mathbf{v}^0(\mathbf{x}) \cdot \nabla \bar{\omega}^0(\mathbf{x})\overset{\eqref{e24}}{=}-\mathbf{v}^0(\mathbf{x})\cdot\nabla\bar{\omega}^0(\mathbf{x})
\end{align*}
hence
\begin{equation}\label{e31}
\partial_t \bar{\mathbf{u}}^{t=0}(\mathbf{x})= \mathbf{K} \ast \partial_t \bar{\omega}^{t=0}(\mathbf{x})=-\mathbf{K}\ast (\mathbf{v}^0 \cdot\nabla\bar{\omega}^0)(\mathbf{x}).
\end{equation}

Let us show the first item in \eqref{e26}. The first assumption in \eqref{e28} is equivalent to
\begin{align}\label{e40}
\bar{\mathbf{u}}^0(\mathbf{x})=(\bar u_1^0(x_2),0).
\end{align}
The velocity of the barycenter is
\begin{align}\label{e27}
\mathbf{G}'(t)=\frac{1}{\Gamma}\int_{\mathbb{T}}\mathbf{x}\partial_t\rho dx=\frac{1}{\Gamma}\int_{\mathbb{T}}\rho(\bar{\mathbf{u}}+\mathbf{v})dx=\frac{1}{\Gamma}\int_{\mathbb{T}}\rho\bar{\mathbf{u}}dx,
\end{align}
where in the last equality we used the antisymmetry of $\mathbf{K}$ to deduce
\begin{align*}
\int\rho\mathbf{v}dx=\int\mathbf{K}(\mathbf{x}, \mathbf{y})\rho(\mathbf{x})\rho(\mathbf{y})dxdy=0.
\end{align*}
Evaluating at $t=0$ and using \eqref{e40}, we deduce $G_2'(0)=0$. 

Regarding the second derivative, let us start be showing that
\begin{align}\label{e32}
G_2''(0)=\Gamma\int_{\mathbb{T}^2}\big(K_2\ast p^\varepsilon(\mathbf{x})\big)^2\partial_2\bar\omega^0(\mathbf{x})dx.
\end{align}
Taking the derivative in \eqref{e27}, we get
\begin{align*}
\mathbf{G}''(t)=\frac{1}{\Gamma} \int \bar{\mathbf{u}}\partial_t \rho  d x +
\frac{1}{\Gamma} \int \rho \partial_t \bar{\mathbf{u}} d x.
\end{align*}
Evaluating at time 0 and using~\eqref{e31}, we have
\begin{align*}
\mathbf{G}''(0) \cdot \hat{\mathbf{x}}_2 &= \frac{\hat{\mathbf{x}}_2}{\Gamma}\cdot \int \rho^0
\big(- \mathbf{K} \ast (\mathbf{v}^0\cdot\nabla \bar{\omega}^0)\big) d x= \frac{\hat{\mathbf{x}}_2}{\Gamma}\cdot \int
(\mathbf{K} \ast \rho^0) (\mathbf{v}^0\cdot \nabla\bar{\omega}^0) d x\\&\overset{\eqref{e29}}{=} 
\frac{\hat{\mathbf{x}}_2}{\Gamma} \cdot \int \mathbf{v}^0(\mathbf{v}^0\cdot \nabla
\bar{\omega}^0)d x\overset{\eqref{e28}}{=}\frac{1}{\Gamma}\int (\mathbf{v}(0,\mathbf{x}) \cdot \hat{\mathbf{x}}_2)^2 \partial_2\bar{\omega}^0(\mathbf{x})dx.
\end{align*}
Using the definitions of $\mathbf{v}$ and $\mathbf{\rho}$ in \eqref{e29} \& \eqref{e30}, we obtain \eqref{e32}.

Let us identify the unit torus with $[-\frac{1}{2},\frac{1}{2}]^2$ and assume without loss of generality that $\mathbf{x}^0=0$. Let us recall that the Biot-Savart kernel on the torus can be written as
\begin{align*}
K_2(\mathbf{x})=\bar{K}_2(\mathbf{x})+h(\mathbf{x})\qquad\mbox{with}\qquad \bar{K}(\mathbf{x})=\frac{\mathbf{x}^\perp}{2\pi|\mathbf{x}|^2},h\in L^\infty(\mathbb{T}^2).
\end{align*}
In order to exploit the better scaling properties of $\bar K$ with respect to $K$, let us show that
\begin{align}\label{e2}
\Big|\int_{\mathbb{T}^2}(K_2\ast p^\varepsilon(\mathbf{x}))^2d x-\int_{\mathbb{T}^2}(\bar{K}_2\ast p^\varepsilon(\mathbf{x}))^2d x\Big|\lesssim 1.
\end{align}
This is a consequence of the following inequalities:
\begin{align*}
&\Big|\int_{\mathbb{T}^2}\big(K_2\ast p^\varepsilon(\mathbf{x})\big)^2 d x-\int_{\mathbb{T}^2}\big(\bar{K}_2\ast p^\varepsilon(\mathbf{x})\big)^2 d x\Big|\le\int_{\mathbb{T}^2}\big|(K_2+\bar{K}_2)\ast p^\varepsilon(\mathbf{x})\big|\big|(K_2-\bar{K}_2)\ast p^\varepsilon(\mathbf{x})\big| d x\\
&\le\big\|(K_2+\bar{K}_2)\ast p^\varepsilon\big\|_{L^1(\mathbb{T}^2)}\big\|h\ast p^\varepsilon\big\|_{L^\infty(\mathbb{T}^2)}\le \big(2\|\bar{K}_2\|_{L^1(\mathbb{T}^2)}+\|h\|_{L^1(\mathbb{T}^2)}\big)\|h\|_{L^\infty(\mathbb{T}^2)}\lesssim 1.
\end{align*}

As a second step, let us prove that
\begin{align}\label{f1}
\Big|\int_{\mathbb{T}^2}(\bar{K}_2\ast p^\varepsilon(\mathbf{x}))^2 d x-\frac{1}{4\pi}\ln\frac{1}{\varepsilon}\Big|\lesssim 1.
\end{align}
The rescaling $\mathbf{x}\mapsto \mathbf{x}':=\varepsilon^{-1}\mathbf{x}$ and the $(-1)$-homogeneity of $\bar{K}_2$ yields the following
\begin{align*}
&\int_{\mathbb{T}^2}\big(\bar{K}_2\ast p^\varepsilon(\mathbf{x})\big)^2 d x\\
&\overset{\eqref{e30}}{=}\int_{\mathbb{T}^2}\Big(\int_{\mathbb{R}^2}\bar{K}_2(\mathbf{x}-\mathbf{y})\frac{1}{\varepsilon^2}p\big(\frac{\mathbf{y}}{\varepsilon}\big) d y\Big)\Big(\int_{\mathbb{R}^2}\bar{K}_2(\mathbf{x}-\mathbf{z})\frac{1}{\varepsilon^2}p\big(\frac{\mathbf{z}}{\varepsilon}\big) d z\Big) d x\\
&=\int_{\frac{1}{\varepsilon}\mathbb{T}^2}\Big(\int_{\mathbb{R}^2}\varepsilon \bar{K}_2(\varepsilon(\mathbf{x}'-\mathbf{y}'))p(\mathbf{y}') d \mathbf{y}'\Big)\Big(\int_{\mathbb{R}^2}\varepsilon \bar{K}_2(\varepsilon(\mathbf{x}'-\mathbf{z}'))p(\mathbf{z}')\big) d z'\Big) d x'\\
&=\int_{\frac{1}{\varepsilon}\mathbb{T}^2}\Big(\int_{\mathbb{R}^2}\bar{K}_2(\mathbf{x}'-\mathbf{y}')p(\mathbf{y}') d \mathbf{y}'\Big)\Big(\int_{\mathbb{R}^2}\bar{K}_2(\mathbf{x}'-\mathbf{z}')p(\mathbf{z}')\big) d z'\Big) d x'\\
&=\int_{\frac{1}{\varepsilon}\mathbb{T}^2}\big(\bar{K}_2\ast p(\mathbf{x})\big)^2 d x.
\end{align*}
Splitting the domain of integration of the last integral, we get
\begin{align*}
\int_{B(0,2)}\big(\bar{K}_2\ast p(\mathbf{x})\big)^2 d x+\int_{\frac{1}{\varepsilon}\mathbb{T}^2\backslash B(0,2)}\big(\bar{K}_2\ast p(\mathbf{x})\big)^2 d x.
\end{align*}
The first term is bounded by $\|\bar{K}_2\|_{L^1(B(0,2))}\|p\|_{L^\infty(B(0,2))}\lesssim 1$, so it is enough to study the second one. Up to an additive error of order 1, we can neglect the convolution since
\begin{align*}
&\Big|\int_{\frac{1}{\varepsilon}\mathbb{T}^2\backslash B(0,2)}\big(\bar{K}_2\ast p(\mathbf{x})\big)^2 d x-\int_{\frac{1}{\varepsilon}\mathbb{T}^2\backslash B(0,2)}\bar{K}_2(\mathbf{x})^2 d x\Big|\\
&\le\int_{\frac{1}{\varepsilon}\mathbb{T}^2\backslash B(0,2)}\Big(\int|\bar{K}_2(\mathbf{x}-\mathbf{y})+\bar{K}_2(\mathbf{x})|p(\mathbf{y}) d y\Big)\Big(\int|\bar{K}_2(\mathbf{x}-\mathbf{y})-\bar{K}_2(\mathbf{x})|p(\mathbf{y}) d y\Big) d x\\
&\lesssim\int_{\frac{1}{\varepsilon}\mathbb{T}^2\backslash B(0,2)}\sup_{\mathbf{y}\in B(0,1)}|\bar{K}_2(\mathbf{x}-\mathbf{y})|\sup_{\mathbf{y}\in B(0,1)}|\nabla \bar{K}_2(\mathbf{x}-\mathbf{y})| d x\\
&\lesssim \int_{\frac{1}{\varepsilon}\mathbb{T}^2\backslash B(0,2)}\sup_{\mathbf{y}\in B(0,1)}\frac{1}{|\mathbf{x}-\mathbf{y}|}\sup_{\mathbf{y}\in B(0,1)}\frac{1}{|\mathbf{x}-\mathbf{y}|^2} d x\lesssim\int_{\mathbb{R}^2\backslash B(0,2)}\frac{1}{|\mathbf{x}|^3}\lesssim 1,
\end{align*}
so that \eqref{f1} will be implied by
\begin{align*}
-1\lesssim \int_{\frac{1}{\varepsilon}\mathbb{T}^2\backslash B(0,2)}\bar{K}_2(\mathbf{x})^2 d x-\frac{1}{4\pi}\ln\frac{1}{\varepsilon}\lesssim 1.
\end{align*}
Using that $\mathbb{T}^2\subset B(0,\frac{1}{\sqrt{2}})$, we deduce the upper bound (the lower bound follows analogously from $\mathbb{T}^2\supset B(0,\frac{1}{2})$):
\begin{align*}
\int_{\frac{1}{\varepsilon}\mathbb{T}^2\backslash B(0,2)}\bar{K}_2(\mathbf{x})^2 d x&\le\int_{B(0,\frac{1}{\sqrt{2}\varepsilon})\backslash B(0,2)}\bar{K}_2(\mathbf{x})^2 d x=\frac{1}{4\pi^2}\int_{B(0,\frac{1}{\sqrt{2}\varepsilon})\backslash B(0,2)}\frac{x_1^2}{|\mathbf{x}|^4} d x\\
&=\frac{1}{8\pi^2}\int_{B(0,\frac{1}{\sqrt{2}\varepsilon})\backslash B(0,2)}\frac{x_1^2+x_2^2}{|\mathbf{x}|^4} d x=\frac{1}{4\pi}\int_2^{\frac{1}{\sqrt{2}\varepsilon}}\frac{1}{\rho} d \rho=\frac{1}{4\pi}\ln\frac{1}{\varepsilon}+O(1).
\end{align*}

From \eqref{e2} and \eqref{f1}, we have proven that
\begin{align*}
\Big|\int_{\mathbb{T}^2}\big(K_2\ast p^\varepsilon(\mathbf{x})\big)^2 d x-\frac{1}{4\pi}\ln\frac{1}{\varepsilon}\Big|\lesssim_{p} 1.
\end{align*}
Starting from this and using \eqref{e32}, then \eqref{e26} is finally established once we prove that
\begin{align*}
\Big|\int_{\mathbb{T}^2}\big(K_2\ast p^\varepsilon(\mathbf{x})\big)^2\partial_2\bar\omega^0(\mathbf{x}) d x-\partial_2\bar{\omega}^0(0)\int_{\mathbb{T}^2}\big(K_2\ast p^\varepsilon(\mathbf{x})\big)^2 d x\Big|\lesssim \|\nabla^2\bar\omega^0\|_\infty,
\end{align*}
which follows from the following sequence of inequalities
\begin{align*}
&\int_{\mathbb{T}^2}\big(K_2\ast p^\varepsilon(\mathbf{x})\big)^2|\partial_2\bar\omega^0(\mathbf{x})-\partial_2\bar\omega^0(0)| d x\lesssim \|\nabla^2\bar\omega^0\|_{L^\infty}\int_{\mathbb{T}^2}(K_2\ast p^\varepsilon(\mathbf{x}))^2|\mathbf{x}| d x\\
&=\|\nabla^2\bar\omega^0\|_{L^\infty}\Big(\int_{B(0,2\varepsilon)}(K_2\ast p^\varepsilon(\mathbf{x}))^2|\mathbf{x}| d x+\int_{\mathbb{T}^2\backslash B(0,2\varepsilon)}(K_2\ast p^\varepsilon(\mathbf{x}))^2|\mathbf{x}| d x\Big)\\
&\le\|\nabla^2\bar\omega^0\|_{L^\infty}\Big(\varepsilon\int_{B(0,2\varepsilon)}(K_2\ast p^\varepsilon(\mathbf{x}))^2 d x+\int_{\mathbb{T}^2\backslash B(0,2\varepsilon)}(\sup_{B(0,\varepsilon)}|K_2(\mathbf{x}-\mathbf{y})|)^2|\mathbf{x}| d x\Big)\\
&\lesssim\|\nabla^2\bar\omega^0\|_{L^\infty}\Big(\varepsilon\ln\frac{1}{\varepsilon}+\int_{\mathbb{T}^2\backslash B(0,2\varepsilon)}\frac{1}{|\mathbf{x}|^2}|\mathbf{x}| d x\Big)\lesssim\|\nabla^2\bar\omega^0\|_{L^\infty}.
\end{align*}
\qed

\begin{rmk}\label{R:1}
From \eqref{e26}, one deduces that the general case (without making
assumptions of the signs of $\Gamma$ and
$\partial_{x_2}\bar{\omega}^0(\mathbf{x}^0)$) is that the map
\[
t \mapsto \text{sign}\big(\Gamma \partial_2\bar{\omega}^0(\mathbf{x}^0)\big) G_2(t)
\]
is strictly increasing. 
\end{rmk}

\begin{rmk}\label{rmk:slope}
Following the same lines of the proof above, the theorem can be extended in the following way. Assume $\bar{\omega}^0(\mathbf{x})$ is a steady solution of the Euler equations (not necessarily depending only on $x_2$). Without loss of generality, assume that $\Gamma > 0$ and that $\nabla \bar{\omega}^0(\mathbf{x}^0) = \partial_2 \bar{\omega}^0(\mathbf{x}^0) \hat{\mathbf{x}}_2$ with $\partial_2 \bar{\omega}^0(\mathbf{x}^0) > 0$ (cf.~Remark~\ref{R:1} for the other cases). Call $\mathbf{G}^\varepsilon(t)$ the position of the barycentre depending on the parameter $\varepsilon$. Then
\[
\lim_{\varepsilon \to 0} G_2^\varepsilon(0)' = 0 \qquad \lim_{\varepsilon \to 0} G_2^\varepsilon(0)'' = +\infty.
\]    
\end{rmk}

\begin{figure}
    \centering
    \includegraphics[width=1\linewidth]{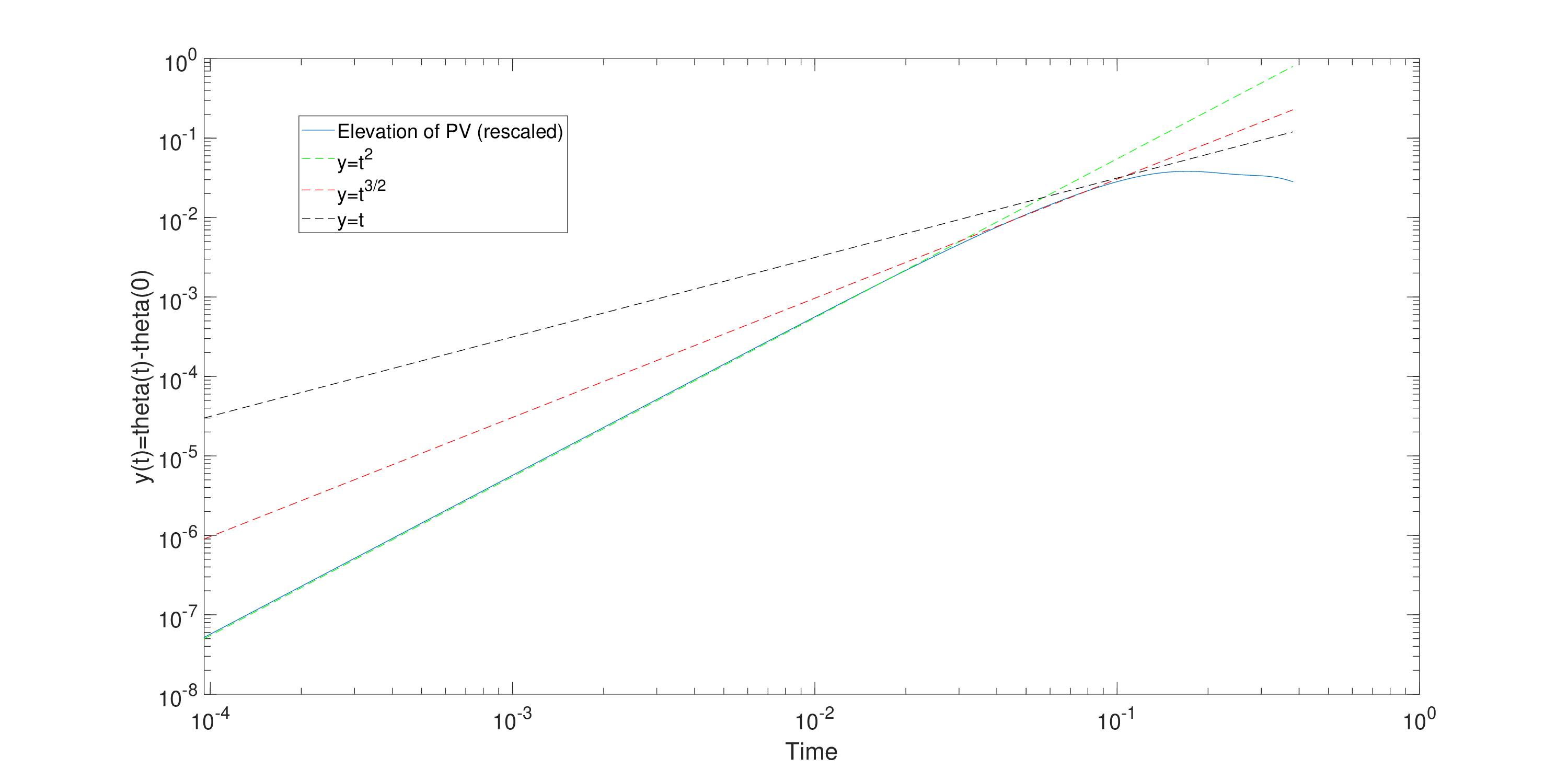}
    \caption{Vertical coordinate trajectory $y=y(t):=\theta(0)-\theta(t)$, plotted in loglog scale, of a point-vortex $x_v$ on a sphere in azimuth-inclination coordinates $(\varphi,\theta)$ with background vorticity $\omega$ equal to the Coriolis vorticity.
    The plot is obtained by a DNS of \eqref{eq:vw_sphere_zeitlin} via the numerical scheme described at the end of Section~\ref{s:vw}, for $N=65$ and time-step $h=10^{-5}$.
    The point-vortex has circulation $\Gamma=0.1$, initial inclination $\theta(0)=\pi-\arccos 0.6$, the Coriolis vorticity is $f = 62\cos \theta$.
    The slope of the curve, after an initial time of approximately $10^{-2}s$, bends towards $\frac{3}{2}$, which would fit with the prediction of Remark~\ref{rmk:slope} of the second derivative being infinity at $t=0$. After $0.1s$ the trajectory of the vortex stops to follow any power law in time, due to the emergence of some more complex dynamics. In the very initial time, the slope of the curve is approximately $2$. This is possibly due to the finite spatial resolution at which \eqref{eq:vw_sphere_zeitlin} is solved. We have empirically observed that increasing the resolution results in a slight shift to the left of the kink between the two slopes.}
    \label{fig:vortex_trajectory}
\end{figure}

\begin{figure}
    \centering
     \includegraphics[width=1.1\linewidth]{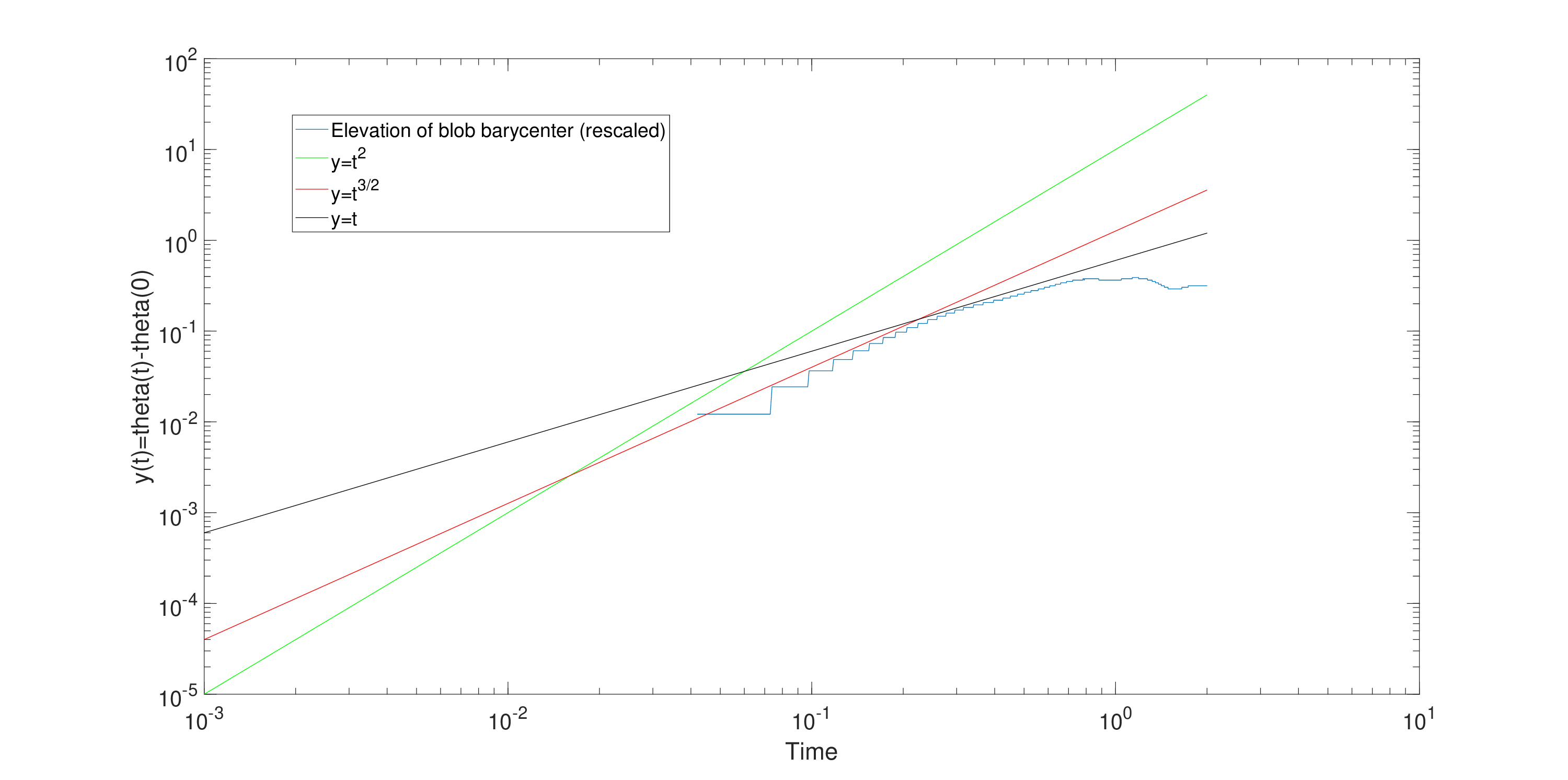}
    \includegraphics[width=1.1\linewidth]{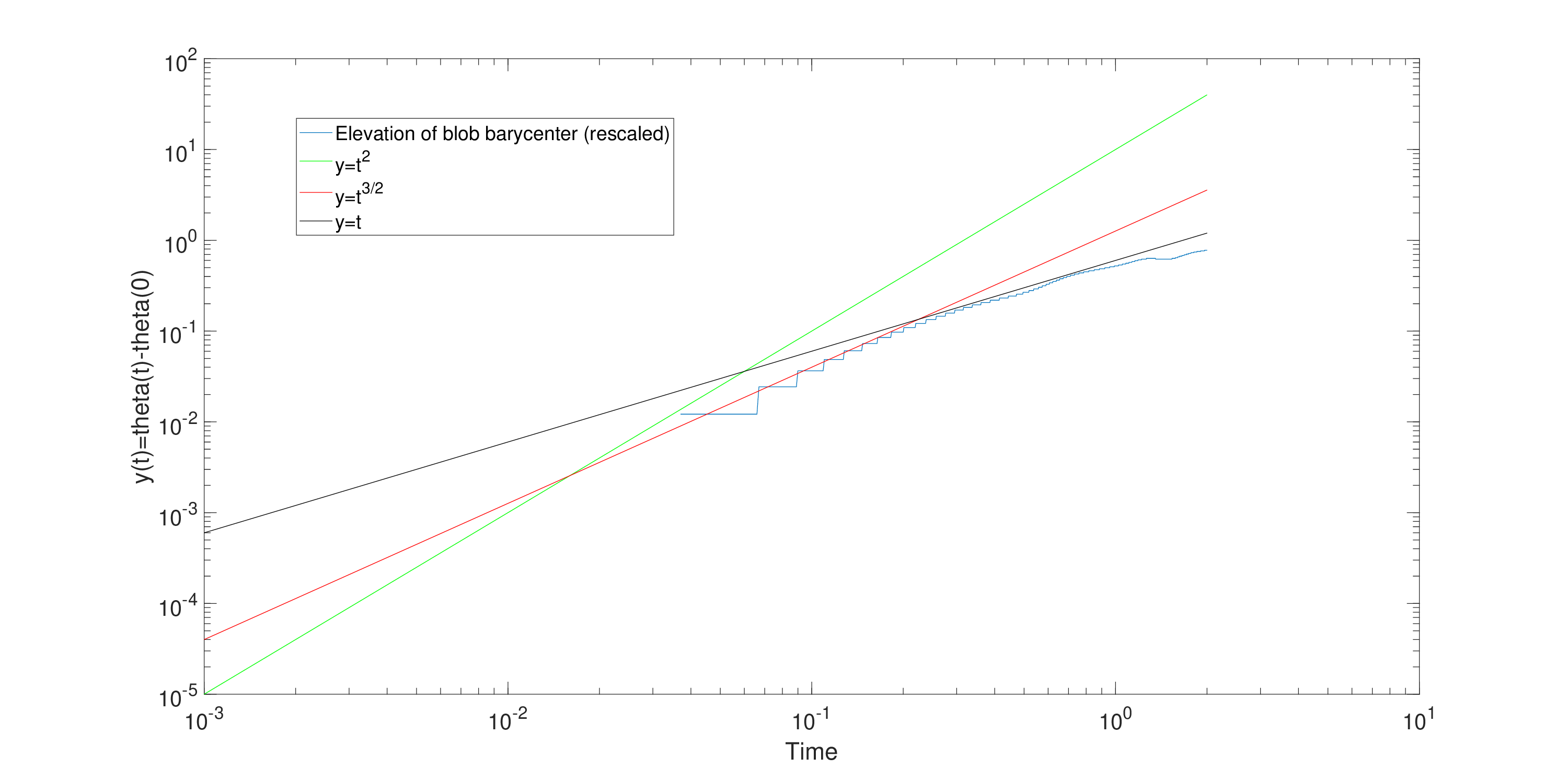}
    \caption{Vertical coordinate trajectory $y=y(t):=\theta(0)-\theta(t)$, plotted in loglog scale, of the barycenter of a blob of vorticity $\omega_b$ of radius $10\varepsilon$ (top) and $\varepsilon$ (bottom), on a sphere in azimuth-inclination coordinates $(\varphi,\theta)$ with background vorticity $\omega_r$ equal to the Coriolis vorticity in the simulation of Figure~\ref{fig:vortex_trajectory}.
       The blob has the same circulation and initial barycenter inclination coordinate of the point-vortex in the simulation of Figure~\ref{fig:vortex_trajectory}.
       The curve $y=y(t)$ is obtained by a DNS of \eqref{eq:vw_sphere_zeitlin} via the numerical scheme described at the end of Section~\ref{s:vw}, for $N=257$, $\varepsilon=1/N$ and time-step $h=10^{-3}$, with no point-vortices ($N_V=0$) and with initial vorticity obtained by the discretization of $\omega=\omega_b+\omega_r$. The slope of the curve, after an initial short time, not captured by the tracking of the blob's barycenter, is around $\frac{3}{2}$, until we see a linear growth.
       We do not see significant differences between the two simulations, denoting a certain stability with respect to the blob size.}
    \label{fig:blob_barycenter_eps}
\end{figure}

Since the second derivative is divergent in the limit $\varepsilon\to0$, it is natural to ask whether the limiting behaviour of the point vortex is of the form
\begin{align*}
G_2(t)\sim t^\alpha\qquad\mbox{for some $\alpha\in(1,2)$.}
\end{align*}
We were unable to theoretically determine it, but we analysed the problem numerically both with the blob-wave dynamics with decreasing radius of the blob $\varepsilon$ (Figure~\ref{fig:blob_barycenter_eps}) and the limiting vortex-wave system (Figure~\ref{fig:vortex_trajectory}), and $\alpha=\frac{3}{2}$ seems to be a fitting exponent. 
The time-scales at which we observe the fitting with the $\frac{3}{2}$ exponent in Figure~\ref{fig:vortex_trajectory} differ by a factor $10$ from Figure~\ref{fig:blob_barycenter_eps}. 
This is conceivable, as we are comparing a point-wise vortex with a blob of positive area.
A precise description of the trajectory of the barycenter of the blob, in the limit of its size vanishing, is beyond the scope of this work and is left for a future study.
We stress that the numerical simulations presented do not validate the theoretical predictions in a strict sense. Nevertheless, we show discrete agreement with the analysis performed in this section. 
We refer to the next section for a more detailed description of the numerical scheme used.

\section{The vortex-wave system}\label{s:vw}

\subsection{The model}

In this section, we recall the results of Marchioro and Pulvirenti about the vortex-wave system (see \cite{MarcPulv}).
\footnote{For our later scopes, we give the results on $\Ss^2$, rather than $\Rr^2$.
We have not seen any crucial difference in changing the reference domain to extend the results of \cite{MarcPulv}.}
Let us consider the incompressible Euler equations on the unit sphere $\mathbb{S}^2\subset\mathbb{R}^3$
\begin{align}\label{e5}
\partial_t\omega+\{\psi,\omega\}=0\qquad\Delta_{\mathbb{S}^2}\psi=\omega,
\end{align}
where $\omega,\psi:\mathbb{S}^2\to\mathbb{R}$ denote respectively the (scalar) vorticity and the stream function and $\{\cdot,\cdot\}$ the Poisson brackets on the sphere. For a classical solution of \eqref{e5}, the velocity of the fluid $\mathbf{u}^t(\mathbf{x})$ defined by
\begin{align}\label{f7}
\mathbf{u}^t(\mathbf{x})=\nabla^\perp_{\mathbb{S}^2}\psi^t(\mathbf{x})\in T_\mathbf{x}\mathbb{S}^2\qquad\mbox{for $t\in[0,\infty)$ and $\mathbf{x}\in\mathbb{S}^2$}
\end{align}
generates a flow of diffeomorphism $\Phi^t:\mathbb{S}^2\to\mathbb{S}^2$ which satisfies
\begin{align*}
\Phi^0(\mathbf{x})=\mathbf{x}\qquad\mbox{and}\qquad\partial_t\Phi^t(\mathbf{x})=\mathbf{u}^t(\Phi^t(\mathbf{x}))\qquad\mbox{for all $t\in[0,\infty)$ and $\mathbf{x}\in\mathbb{S}^2$,}
\end{align*}
and so that the transport formula $\omega^t(\mathbf{x})=\omega^0\big((\Phi^t)^{-1}(\mathbf{x})\big)$ holds.

We will consider an initial condition $\omega^0$ which is a (possibly signed) measure of the form
\begin{align*}
\omega^0(d\mathbf{x})=\bar{\omega}^0(\mathbf{x})d\mathbf{x}+ \Gamma\delta_{\mathbf{x}^0}(d\mathbf{x}),
\end{align*}
where $\bar\omega^0\in L^\infty(\mathbb{S}^2)$ denotes a large-scale background vorticity and $\Gamma\delta_{\mathbf{x}^0}$ a point-like vortex located at $\mathbf{x}^0\in\mathbb{S}^2$ of circulation $\Gamma$.
In \cite{MarcPulv}, denoting, respectively, by $\bar\omega^t$ and $\mathbf{x}_v(t)\in\mathbb{S}^2$ the background vorticity and the position of the vortex at time $t$, the time-dependent measure
\begin{align*}
\omega^t(d\mathbf{x})=\bar{\omega}^t(\mathbf{x}) d \mathbf{x}+\Gamma\delta_{\mathbf{x}_v(t)}( d \mathbf{x})
\end{align*}
is said to satisfy the Euler equations in a weak form if $\bar{\omega}\in L^\infty([0, +\infty)\times\mathbb{S}^2)$, $\mathbf{x}_v \in C^1([0, +\infty), \Ss^2)$ and the following system of equations holds
\begin{equation}\label{eq:vw}
\begin{cases}
\partial_t\Phi^t(\mathbf{x})=\bar{\mathbf{u}}^t\big(\Phi^t(\mathbf{x})\big)+\Gamma \mathbf{K}\big(\Phi^t(\mathbf{x}),\mathbf{x}_v(t)\big),\quad\Phi^0(\mathbf{x})=\mathbf{x}, \;\mathbf{x}\neq\mathbf{x}^0\\
\displaystyle\frac{ d \mathbf{x}_v}{ d t}(t)=\bar{\mathbf{u}}^t\big(\mathbf{x}_v(t)\big),\quad\mathbf{x}_v(0)=\mathbf{x}^0\\
\bar{\mathbf{u}}^t(\mathbf{x})=\displaystyle\int_{\mathbb{S}^2} \mathbf{K}(\mathbf{x},\mathbf{y})\bar{\omega}^t(\mathbf{y}) d y \\
\bar{\omega}^t(d\mathbf{x})=\bar{\omega}^0\big((\Phi^t)^{-1}(\mathbf{x})\big)
\end{cases}
\end{equation}
Here $\bar{\mathbf{u}}$ is the velocity field given by $\bar{\omega}$, the bounded part of the vorticity. In this section, the function $\mathbf{K}=\mathbf{K}_{\mathbb{S}^2}$ is the Biot-Savart kernel on $\Ss^2$
\begin{align*}
\mathbf{K}(\mathbf{x},\mathbf{y}):=\nabla^\perp_{\mathbb{S}^2}\Delta_{\mathbb{S}^2}^{-1}\delta_\mathbf{y}(\mathbf{x})=\frac{1}{2\pi}\frac{(\mathbf{x}-\mathbf{y})\times\mathbf{y}}{|\mathbf{x}-\mathbf{y}|^2}.
\end{align*}
The point vortex $\mathbf{x}_v$ follows only the velocity field $\bar{\mathbf{u}}$, discarding the self-interaction, while $\bar{\omega}$ evolves transported by the sum of the divergent velocity field of the vortex and its velocity field $\bar{\mathbf{u}}$. 

Some results from \cite{MarcPulv} are summarised in the following theorem.

\begin{thm}
Given initial conditions $\bar{\omega}^0 \in L^\infty(\Ss^2)$, $\mathbf{x}^0 \in \Ss^2$ and $\Gamma\in\Rr\backslash\{0\}$, there exists a global in time solution of system \eqref{eq:vw}. Under the additional assumption that $\bar{\omega}^0$ is constant in a neighbourhood of $\mathbf{x}^0$, the solution is unique and it is the weak-* limit (in the sense of finite measures) as $\varepsilon\downarrow0$ of the solutions starting from the bounded initial conditions
\begin{align*}
\omega^{0,\varepsilon}(\mathbf{x}):=\bar{\omega}^0(\mathbf{x})+ \frac{\Gamma}{\pi \varepsilon^2}\chi_{B(\mathbf{x}^0, \varepsilon)}
\end{align*}
where $\chi_{B(\mathbf{x}^0,\varepsilon)}$ denotes the characteristic function of the ball $B(\mathbf{x}^0,\varepsilon)$ centered at $\mathbf{x}^0$ of radius $\varepsilon$.
\end{thm}

Due to the divergence of the vector field of the vortex, the case when $\bar{\omega}^0$ is not locally constant is less understood. In the smooth case, $\bar{\omega}^0$ is not locally constant in particular if $\nabla \bar{\omega}^0(\mathbf{x}^0) \neq 0$. Under this assumption, we provide numerical simulations (Section~\ref{num}) and a rigorous proof (Section~\ref{blob}) that the vortex perceives a very strong acceleration proportional to $\Gamma\nabla\bar{\omega}(\mathbf{x}^0)$. 

\subsection{Direct numerical simulations}\label{num}

In this section, we specifiy the vortex-wave equations \eqref{eq:vw} on a rotating 2-sphere $\Ss^2\subset\Rr^3$.
These equations were first introduced by Bogomolov \cite{Bog1977} and more recently studied by Laurent--Polz in \cite{Lau2005}, in the context of geometric mechanics.
Formally, we set the equations in the following notations:
\begin{itemize}
\item Position of the point vortices: $\mathbf{x}_v^i:[0,T]\rightarrow\Ss^2$, for $i=1,...,N_V$. 
\item Background vorticity: $\bar\omega:[0,T]\times \Ss^2\rightarrow\Rr$.
\item \textit{Coriolis vorticity}: $f=f(\mathbf{x}):=\mathbf{\Omega}\cdot\mathbf{x}$, for $\mathbf{\Omega}\in\Rr^3$ angular velocity of the rotating sphere.
\end{itemize}

Then the vortex-wave equations on a rotating sphere are given by:

\begin{equation}\label{eq:vw_sphere}
\begin{array}{cc}
  &\dot{\mathbf{x}}_v^i = \dfrac{1}{4\pi}\displaystyle\sum_{\substack{j=1,..,N_V \\ i\neq j}}\Gamma_j\dfrac{\mathbf{x}_v^i\times\mathbf{x}_v^j}{1-\mathbf{x}_v^i\cdot\mathbf{x}_v^j}+\mathbf{x}_v^i\times\nabla\Delta_{\mathbb{S}^2}^{-1}(\bar\omega-f)(\mathbf{x}_v^i) \hspace{.5cm} \mbox{for } i=1,...,N_V\\
    &\dot{\bar{\omega}} =\lbrace\bar\omega , \Delta_{\mathbb{S}^2}^{-1}(\bar\omega-f) + \displaystyle\sum_{i=1,\ldots,N_V}\Gamma_i\log(1-\mathbf{x}_v^i\cdot\mathbf{x})\rbrace.
\end{array}
\end{equation}
The Poisson bracket of two functions $f,g$ is defined as $\lbrace f,g\rbrace:=\textbf{x}\cdot(\nabla f\times\nabla g).$
We remark that the Laplace-Beltrami operator on the sphere $\Delta_{\mathbb{S}^2}$ is invertible on the space of functions with zero mean. 
Equations \eqref{eq:vw_sphere} are a infinite dimensional Hamiltonian system (more precisely Lie--Poisson system) with energy
\[
H=-\frac{1}{2}\sum_{i\neq j} \Gamma_i\Gamma_j \log(1-\mathbf{x}_v^i\cdot \mathbf{x}_v^j) - \sum_{i}\Gamma_i(\bar\omega-f)(\mathbf{x}_v^i) - \frac{1}{2}\int_{\Ss^2}\Delta_{\mathbb{S}^2}^{-1}(\bar\omega-f)(\bar\omega-f).
\]
Equations \eqref{eq:vw_sphere} also have additional Casimir invariants given by $C_f=\int_{\Ss^2}f(\bar\omega)$, for any real function $f$ and the angular momentum $M=\int_{\Ss^2}(\sum_i \Gamma_i\delta_{\mathbf{x}_v^i}+\bar\omega)\textbf{n}dS\cdot\mathbf{\Omega}$, for $\mathbf{\Omega}\neq 0$.

A spatial discretization which retains many conservation laws is known for the 2D Euler equations as the Zeitlin model \cite{Zei2004}.
In the following discussions, we use the conventions of \cite{ModViv2025}. Starting from equations \eqref{eq:vw_sphere}, we obtain the vortex-wave--Zeitlin model.
Given the expansion of the vorticity field in spherical harmonics $Y_{lm}$, we take the truncated expansion for $l\leq N-1,m=-l,...,l$.
Then, for any $N\geq 1$, we have an explicit vector space isomorphism between the one spanned by classical spherical harmonics $\lbrace Y_{lm}\rbrace_{\substack{l=1,..,N-1 \\ m=-l,...,l}}$ and a $\SU(N)$ (skew-Hermitian matrix with zero trace), spanned by the matrices $\lbrace T_{lm}\rbrace_{\substack{l=1,..,N-1 \\ m=-l,...,l}}$.
In other words, if $\bar{\omega}=\sum_{l=1}^\infty\sum_{m=-l}^l\bar{\omega}^{lm}Y_{lm}$, then $W=\sum_{l=1}^{N-1}\sum_{m=-l}^l\bar{\omega}^{lm}T_{lm}$.
In these notations, the vortex-wave--Zeitlin is given by: 
\begin{equation}\label{eq:vw_sphere_zeitlin}
\begin{array}{cc}
&\dot{\mathbf{x}}_v^i  = \dfrac{1}{4\pi}\sum_{\substack{j=1,..,N_V \\ i\neq j}}\Gamma_i\dfrac{\mathbf{x}_v^i\times\mathbf{x}_v^j}{1-\mathbf{x}_v^i\cdot\mathbf{x}_v^j}-\sum_{\substack{l=1,..,N-1\\m=-l,...,l}}\dfrac{\mathbf{x}_v^i\times\nabla Y_{lm}(\mathbf{x}_v^i)}{-l(l+1)}(W-F)^{lm} \hspace{.3cm} \mbox{ for } i=1,...,N_V\\
&\\
&\dot{W} = - [W , \Delta_N^{-1}\big(W+ \sum_{\substack{i=1,..,N_V \\ l=1,..,N-1 \\ m=-l,...,l}}\Gamma_i Y_{lm}(\mathbf{x}_v^i)T_{lm}- F\big)]_N,
\end{array}
\end{equation}
where $\Delta_N$ is the discrete Laplacian, $[A,B]_N=\frac{\sqrt{N^2-1}}{2}(AB-BA)$ and $F=\vert\mathbf{\Omega}\vert i T_{10}$. The Hamiltonian is:
\begin{equation}
\begin{array}{cc}
H = &-\dfrac{1}{8\pi}\sum_{\substack{i,j=1,..,N_V \\ i\neq j}}\Gamma_i\Gamma_j\log(1-\mathbf{x}_i\cdot\mathbf{x}_j) +\\
&- \sum_{\substack{i=1,..,N_V \\ l=1,..,N-1 \\ m=-l,...,l}}\Gamma_i \dfrac{Y_{lm}(\mathbf{x}_i)}{-l(l+1)}(W-F)^{lm} +\\
&- \dfrac{1}{2}\Tr(\Delta_N^{-1}(W-F)^*(W-F)).
\end{array}
\end{equation}
Equations \eqref{eq:vw_sphere_zeitlin} have the following independent first integrals:
\begin{equation}\label{first_integrals_quant}
\begin{array}{cc}
&C_n(W) = \Tr(W^n), \mbox{ for } n=2,...,N-1
\\
&\\
&M = \left(\sum_{i=1}^{N_V}\Gamma_i\mathbf{x}_v^i + \left[\begin{matrix}
\Tr(W^*T_x)\\
\Tr(W^*T_y) \\
\Tr(W^*T_z) \\
\end{matrix}\right]\right)\cdot\mathbf{\Omega},
\end{array}
\end{equation}
where $T_x=\frac{iT_{1-1}-iT_{1-1}}{\sqrt{2}},T_y=\frac{T_{1-1}+T_{1-1}}{\sqrt{2}},T_z=iT_{10}$.

\begin{figure}
    \centering
\includegraphics[width=1\linewidth]{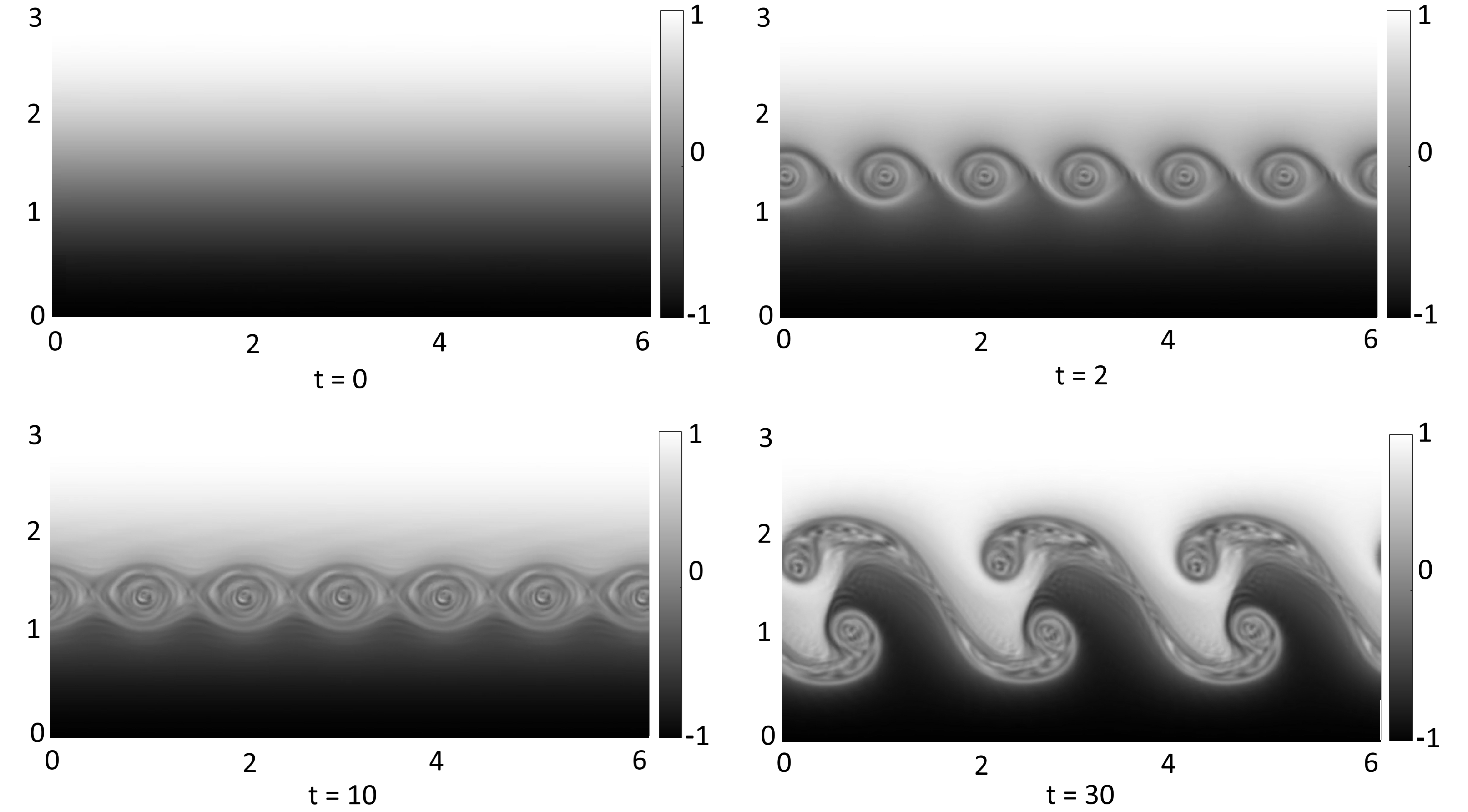}
    \caption{From the top-left to the bottom-right, time snapshots of the evolution of the vorticity field $\omega$ on a 2-sphere, interacting with six equal equatorial point vortices.
    The plots, in azimuth-inclination coordinates $(\varphi,\theta)$, are obtained by a DNS of \eqref{eq:vw_sphere_zeitlin} for $N=501$. The initial (unstable) equilibrium of 6 equatorial point-vortices placed at the vertices of a regular equatorial hexagon, all with the same vorticity $\Gamma>0$, surrounded by an initial smooth field equal to the Coriolis vorticity $\omega^0=2\Omega\cos\theta$, for $\Omega=-1/2$, evolves into a coupled vortex-wave dynamics.
    Here, the self-interaction among the point-vortices is crucial to develop this kind of behaviour. The interaction with the background shear flow acts only as a perturbation of the initial equilibrium configuration of point-vortices.}
    \label{fig:DNS_vortex_wave}
\end{figure}

A conservative time-discretization of equations \eqref{eq:vw_sphere_zeitlin} can be obtained by extending the method used in \cite{ModViv2020}\footnote{The code used to perform numerical simulations of the Zeitlin model can be found \url{https://github.com/klasmodin/quflow}. The code for the vortex-wave Zeitlin model is not available yet.} to the coupled vortex-wave system. 
Therefore, we obtain a Lie--Poisson integrator for \eqref{eq:vw_sphere_zeitlin} that exactly conserves the integrals $C_n$ and $M$ and nearly conserves the Hamiltonian $H$.
In Figure~\ref{fig:DNS_vortex_wave}, we illustrate an example of a DNS of \eqref{eq:vw_sphere_zeitlin}, solved with the Lie--Poisson integrator developed in \cite{ModViv2020}, in which 6 identical point-vortices are placed in the vertices of a regular equatorial hexagon and a background smooth vorticity equal to the Coriolis one.

\section{Quasi-two-dimensional flows}

Most real systems where, in a first approximation a 2D model is investigated,
are indeed quasi-two-dimensional. The deviation from 2D may be relevant and
the consequences very important. We address to \cite{Alexakis} for a review of
problems and results on this topic.

The question we discuss in this section is whether our first-stage result of
motion along the gradient extends to thin vortex tubes in 3D\ shear flows,
with approximate symmetries that make the system close to 2D. 

The first difficulty in order to develop a theory is the concept of
\textit{vortex tube} and its evolution. See \cite{Fefferman} as an instance of
the difficulty to handle this topic. Just to mention a technical problem we
encountered when trying to extend our 2D result, the concept of center of mass
$\mathbf{G}\left(  t\right)  $ introduced above is not very clear. Can we define a
curve inside the vortex tube which replaces the center of mass?

What is easy to define is the concept of \textit{vortex filament}: a
time-dependent parametrized curve of the form%
\[
\mathbf{\gamma}^{t}\left(  r\right)  \in\mathbb{R}^{3},\qquad r\in\mathbb{R}%
\]
where $t$ is time and $r$ is the parameter. Given a circulation parameter
$\Gamma$, we may define the associated vorticity field as the vector valued
distribution \cite{Bessaih}%
\[
\mathbf{\omega}^{\prime}\left(  t,\mathbf{x}\right)  =\Gamma\int_{\mathbb{R}%
}\delta\left(  \mathbf{x}-\mathbf{\gamma}^{t}\left(  r\right)  \right)
\partial_{r}\mathbf{\gamma}^{t}\left(  r\right)  dr
\]
(we use the notation $\mathbf{\omega}^{\prime}$ as if it were a fluctuation, a
perturbation of a field). We may also create at time $t=0$ a smoother version
by a convolution%
\[
\mathbf{\omega}_{\varepsilon}^{\prime}\left(  0,\mathbf{x}\right)  =\Gamma
\int_{\mathbb{R}}p^\varepsilon\left(  \mathbf{x}-\mathbf{\gamma}%
^{0}\left(  r\right)  \right)  \partial_{r}\mathbf{\gamma}^{0}\left(
r\right)  dr
\]
where $p^\varepsilon\left(  \mathbf{x}\right)  :=\varepsilon^{-3}p\left(
\varepsilon^{-1}\mathbf{x}\right)  $ and $p$ is a smooth compact support
function of integral 1. However, opposite to the fact that the filament remains a curve for
larger times, what happens to $\mathbf{\omega}_{\varepsilon}^{\prime}\left(
0,\mathbf{x}\right)  $ is less easy to describe. Also an initially circular
blob in 2D is subject to strong deformations as time evolves, but at least we
may investigate its baricenter. In the case of a vortex tube, this is more
difficult. 

One is therefore tempted to work with only the vortex filament, and
investigate its evolution in a background shear flow. However, as explained in
\cite{berselli}, \cite{Bessaih} and other works, the dynamics of a vortex
filament is too singular and the only way to study it is by introducing a
regularization (a cut-off) of the Biot-Savart kernel.

In the next two subsections, we proceed in this way, by mollification of the
Biot-Savart kernel. We start in Section \ref{section 3D vortex filament}, working on the torus $\mathbb{T}^3$, with a slightly tilted vortex filament. In Section~\ref{S:thin}, we derive the result for a thin three-dimensional domain $\mathbb{T}^3_\delta$.

\subsection{A vortex filament in a smoothed Biot-Savart
set-up\label{section 3D vortex filament}}

Given a smooth $\alpha:\mathbb{T}\to\mathbb{T}$, let us consider an initial vortex filament in the three dimensional torus $\mathbb{T}^3$
\begin{align}
\gamma^0:\mathbb{T}\to\mathbb{T}^3,\qquad \gamma^0(r)=\alpha(r)\hat{\mathbf{x}}_2+r\hat{\mathbf{x}}_3\label{e7},
\end{align}
whose maximal slope with respect to the vertical line is described by the non-dimensional parameter
\begin{align}\label{e41}
\eta:=\|\alpha'(r)\|_\infty.
\end{align}
We will work in the regime $\eta\ll1$, namely the vortex filament is just slightly tilted with respect to the vertical axis.

The velocity and the vorticity of the background shear flow are assumed to be of the form
\begin{align}\label{e3}
\bar{\mathbf{u}}^0(\mathbf{x}):=(\bar{u}^0_1(x_2),0,0)\qquad\mbox{and}\qquad\bar\omega^0(\mathbf{x}):=(0,0,\bar\omega^0_3(x_2)),
\end{align}
so that from the relation $\bar\omega=\text{curl}~\bar u$, we learn that
\begin{align}\label{e42}
\bar\omega^0_3(x_2)=-\frac{d}{dx_2}\bar u^0_1(x_2).
\end{align}

Let us denote by $\mathbf{k}$ the Biot-Savart kernel on the torus $\mathbb{T}^3$ (identified in the sequel with $[-\frac{1}{2},\frac{1}{2}]^3$) and recall the following property 
\begin{align*}
\mathbf{k}(\mathbf{x})=\bar{\mathbf{k}}(\mathbf{x})+\mathbf{h}(\mathbf{x}),\quad\mbox{with}~\bar{\mathbf{k}}(\mathbf{x})=\frac{1}{4\pi}\frac{\mathbf{x}}{|\mathbf{x}|^3}~\mbox{and}~\mathbf{h}\in C^\infty\big([-\frac{1}{2},\frac{1}{2}]^3,\mathbb{R}^3\big).
\end{align*}
We will use the regularized version $\mathbf{k}^\varepsilon$ of $\mathbf{k}$ defined by
\begin{align*}
\mathbf{k}^\varepsilon(\mathbf{x}):=\frac{1}{4\pi}\frac{\mathbf{x}}{(|\mathbf{x}|^2+\varepsilon^2)^{\frac{3}{2}}}+\mathbf{h}^\varepsilon(\mathbf{x}).
\end{align*}
Here $\mathbf{h}_\varepsilon$ is a slight perturbation of $\mathbf{h}$ which makes the kernel $\mathbf{k}^\varepsilon$ periodic. We will only use that is satisfies uniform (in $\varepsilon$) $L^\infty$ bounds.

The dynamics is the following
\begin{align}
&\partial_t\bar\omega+(\mathbf{u}+\mathbf{v})\cdot\nabla\bar\omega-\bar\omega\cdot\nabla(\mathbf{u}+\mathbf{v})=0,\label{e44}\\
&\bar{\mathbf{u}}^t(\mathbf{x})=\int_{\mathbb{T}^3}\mathbf{k}^\varepsilon(\mathbf{x}-\mathbf{y})\times\bar\omega^t(\mathbf{y})dy,\label{e45}\\
&\mathbf{v}^t(\mathbf{x})=\Gamma\int_{\mathbb{T}}\mathbf{k}^\varepsilon(\mathbf{x}-\gamma^t(r))\times\partial_r\gamma^t(r)dr,\label{e6}\\
&\partial_t\gamma^t(r)=\bar{\mathbf{u}}^t(\gamma^t(r))+\mathbf{v}^t(\gamma^t(r))\label{e1}.
\end{align}

The main result of this section is the following.

\begin{thm}\label{T:2}
Assume that $\partial_2\bar{\omega}_3^0(\gamma^0(r_0))>0$ for some $r_0\in\mathbb{T}$. For $\bar\omega^0$ fixed, in the limit
\begin{align}\label{e43}
\varepsilon\ll1\qquad\mbox{and}\qquad\eta\ll\varepsilon^3\frac{\partial_2\bar\omega_3^0(\gamma^0(r_0))}{\Gamma},
\end{align}
we have that
\begin{align*}
\partial_t\gamma_2^{t=0}(r_0)=0\qquad\mbox{and}\qquad
\partial_t^2\gamma^{t=0}_2(r_0)\sim\frac{\Gamma\partial_2\bar\omega^0_3(\gamma^0(r_0))}{4\pi}\ln\frac{1}{\varepsilon}.
\end{align*}
\end{thm}

To do the proof, we need some technical lemmas. The first two give estimates on the regularized Biot-Savart kernel.

\begin{lemma}\label{L:1}
\begin{align}
&\int_{\mathbb{T}}\big|\partial_ik^\varepsilon_j(\gamma^0(r_0)-\gamma^0(r))\big|dr\lesssim\varepsilon^{-2},\label{e16a}\\
&\int_{\mathbb{T}}\big|\partial_ik_j^\varepsilon(\gamma^0(r_0)-\gamma^0(r))\big||r_0-r|dr\lesssim\varepsilon^{-1},\label{e16b}\\
&\int_{\mathbb{T}}\big|k_3^\varepsilon(\gamma^0(r_0)-\gamma^0(r))\big|dr\lesssim\varepsilon^{-1}.\label{e16c}
\end{align}
\end{lemma}

{\sc Proof.} Let us show \eqref{e16a}, since the other two are analogous. From the explicit expression
\begin{align*}
\partial_ik_j^\varepsilon(\mathbf{x})=\delta_{ij}(|\mathbf{x}|^2+\varepsilon^2)^{-\frac{3}{2}}-3x_ix_j(|\mathbf{x}|^2+\varepsilon^2)^{-\frac{5}{2}}+\partial_ih^\varepsilon_j
\end{align*}
we learn that $|\partial_ik_j^\varepsilon|(\mathbf{x})\lesssim(|\mathbf{x}|^2+\varepsilon^2)^{-\frac{3}{2}}$. This implies
\begin{align}\label{e23}
\int_{\mathbb{T}}\big|\partial_ik_j^\varepsilon(\gamma^0(r^0)-\gamma^0(r))\big|dr&\lesssim\int_{\mathbb{T}}(|\gamma^0(r_0)-\gamma^0(r)|^2+\varepsilon^2)^{-\frac{3}{2}}dr\le \int_{\mathbb{T}}\big((r-r_0)^2+\varepsilon^2\big)^{-\frac{3}{2}}dr\nonumber\\
&\lesssim\int_{|r-r_0|\ge\varepsilon}(r-r_0)^{-3}dr+\int_{|r-r_0|\le\varepsilon}\varepsilon^{-3}dr\lesssim\varepsilon^{-2}.
\end{align}
\qed

\begin{lemma}\label{L:2}
There exists a function $\phi:[0,\infty)\to[0,\infty)$ and a constant $C>0$ with the properties
\begin{align}\label{e22}
\mbox{$\phi$ is increasing,}\qquad\phi(t)\le Ct,\qquad\lim_{t\to\infty}\phi(t)=1,
\end{align}
such that the following expansion holds
\begin{align*}
k_1^\varepsilon\ast k_1^\varepsilon(\mathbf{x})=-\frac{1}{4\pi|\mathbf{x}|}\phi\big(\varepsilon^{-1}|\mathbf{x}|\big)+O(1)\qquad\mbox{for all}~\mathbf{x}=(0,x_2,x_3).
\end{align*}
\end{lemma}

{\sc Proof.} Up to losing an error of order 1, we can prove the expansion for $\bar k_1^\epsilon$ (in place of $k_1^\epsilon$), for which we have
\begin{align*}
\bar k_1^\varepsilon\ast\bar k_1^\varepsilon(\mathbf{x})=\frac{1}{16\pi^2}\int_{\mathbb{T}^3}\frac{x_1-y_1}{(|\mathbf{x}-\mathbf{y}|^2+\varepsilon^2)^{\frac{3}{2}}}\frac{y_1}{(|\mathbf{y}|^2+\varepsilon^2)^{\frac{3}{2}}}dy.
\end{align*}
With the identification $\mathbb{T}^3\simeq[-\frac{1}{2},\frac{1}{2}]^3$ and for $\mathbf{x}\in B(0,\frac{1}{4})$, we can extend the integral to the all space $\mathbb{R}^3$, by making an error of order one, since
\begin{align*}
\Big|\int_{\mathbb{R}^3\backslash[-\frac{1}{2},\frac{1}{2}]^3}\frac{x_1-y_1}{(|\mathbf{x}-\mathbf{y}|^2+\varepsilon^2)^{\frac{3}{2}}}\frac{y_1}{(|\mathbf{y}|^2+\varepsilon^2)^{\frac{3}{2}}}dy\Big|\lesssim\int_{\mathbb{R}^3\backslash[-\frac{1}{2},\frac{1}{2}]^3}\frac{1}{|\mathbf{y}|^4}dy\lesssim 1
\end{align*}
where in the first inequality we compared with $\varepsilon=0$ and used that
\begin{align*}
|\mathbf{x}-\mathbf{y}|\ge\frac{1}{2}|\mathbf{y}|\qquad\mbox{for all}~\mathbf{x}\in B(0,\frac{1}{4}),\mathbf{y}\in\mathbb{R}^3\backslash[-\frac{1}{2},\frac{1}{2}]^3.
\end{align*}

If $x_1=\mathbf{x}\cdot\hat{\mathbf{x}}_1=0$, then
\begin{align*}
&\bar k_1^\varepsilon\ast\bar k_1^\varepsilon(\mathbf{x})=-\frac{1}{16\pi^2}\int_{\mathbb{R}^3}\frac{y_1^2}{(|\mathbf{x}-\mathbf{y}|^2+\varepsilon^2)^{\frac{3}{2}}(|\mathbf{y}|^2+\varepsilon^2)^{\frac{3}{2}}}dy+O(1)\\
&=-\frac{1}{16\pi^2}\int_{\mathbb{R}^3}\frac{y_1^2}{(|\mathbf{y}-|\mathbf{x}|\hat{\mathbf{x}}_3|^2+\varepsilon^2)^{\frac{3}{2}}(|\mathbf{y}|^2+\varepsilon^2)^{\frac{3}{2}}}dy+O(1)\qquad\mbox{(Rotation in the $\hat{\mathbf{x}}_2,\hat{\mathbf{x}}_3$ plane.)}\\
&=-\frac{1}{16\pi^2|\mathbf{x}|}\int_{\mathbb{R}^3}\frac{y_1^2}{(|\mathbf{y}-\hat{\mathbf{x}}_3|^2+(\frac{\varepsilon}{|\mathbf{x}|})^2)^{\frac{3}{2}}(|\mathbf{y}|^2+(\frac{\varepsilon}{|\mathbf{x}|})^2)^{\frac{3}{2}}}dy+O(1)\qquad\mbox{(Change of variables $\mathbf{y}\mapsto|\mathbf{x}|\mathbf{y}$.)}\\
&=-\frac{1}{4\pi|\mathbf{x}|}\phi\big(\varepsilon^{-1}|\mathbf{x}|\big)+O(1),
\end{align*}
where we introduced
\begin{align*}
\phi(t):=\frac{1}{4\pi}\int_{\mathbb{R}^3}\frac{y_1^2}{(|\mathbf{y}-\hat{\mathbf{x}}_3|^2+t^{-2})^{\frac{3}{2}}(|\mathbf{y}|^2+t^{-2})^{\frac{3}{2}}}dy.
\end{align*}

It remains to show \eqref{e22}. The monotonicity follows from the definition, while the limit at infinity follows from the explicit integral
\begin{align*}
\int_{\mathbb{R}^3}\frac{y_1^2}{|\mathbf{y}-\hat{\mathbf{x}}_3|^3|\mathbf{y}|^3}dy=4\pi.
\end{align*}
Regarding the linear bound in $t$, it is enough to prove it for $t<\frac{1}{2}$, for which we have
\begin{align*}
\phi(t)&\lesssim\int_{B(0,t^{-1})}\frac{y_1^2}{(|\mathbf{y}-\hat{\mathbf{x}}_3|^2+t^{-2})^{\frac{3}{2}}(|\mathbf{y}|^2+t^{-2})^{\frac{3}{2}}}dy+\int_{B(0,t^{-1})^c}\frac{y_1^2}{(|\mathbf{y}-\hat{\mathbf{x}}_3|^2+t^{-2})^{\frac{3}{2}}(|\mathbf{y}|^2+t^{-2})^{\frac{3}{2}}}dy\\
&\lesssim t^6\int_{B(0,t^{-1})}|\mathbf{y}|^2dy+\int_{B(0,t^{-1})^c}|\mathbf{y}|^{-4}dy\lesssim t.
\end{align*}
\qed

Now comes a lemma which exploits two quasi-two-dimensionality to derive estimates on the velocity field generated by the vortex filament.

\begin{lemma}
\begin{align}
&\sup_{r\in\mathbb{T}}|v_1^0(\gamma^0(r))|\lesssim\Gamma\eta\varepsilon^{-1}\label{e18a}\\
&\sup_{\mathbf{x}\in\mathbb{T}^3}|\partial_3\mathbf{v}^0(\mathbf{x})|\lesssim\Gamma\eta\varepsilon^{-2}\label{e18b}.
\end{align}
\end{lemma}

{\sc Proof.} From \eqref{e7} and \eqref{e6}, we have that
\begin{align*}
v^0_1(\gamma^0(r))=\Gamma\int_{\mathbb{T}}k_2^\varepsilon(\gamma^0(r)-\gamma^0(r'))dr'-\Gamma\int_{\mathbb{T}}\alpha'(r')k_3^\varepsilon(\gamma^0(r)-\gamma^0(r'))dr'.
\end{align*}
The second term is estimated by \eqref{e16c} and definition \eqref{e41}. For the second term, we notice that
\begin{align*}
\big|k_2^\varepsilon(\gamma^0(r)-\gamma^0(r'))\big|\lesssim\frac{|\gamma_2(r)-\gamma_2(r')|}{(|r-r'|^2+\varepsilon^2)^{\frac{3}{2}}}\lesssim\eta(|r-r'|^2+\varepsilon^2)^{-1},
\end{align*}
and apply the splitting $\mathbb{T}=\{|r-r'|\ge\varepsilon\}\cup\{|r-r'|<\varepsilon\}$ as in \eqref{e23}. This proves \eqref{e18a}.

Regarding \eqref{e18b}, we have
\begin{align*}
\partial_3\mathbf{v}(\mathbf{x})&=\Gamma\int_{\mathbb{T}}\partial_3\mathbf{k}^\varepsilon(\mathbf{x}-\gamma^0(r))\times\partial_r\gamma^0(r)dr=\\
&=\Gamma\int_{\mathbb{T}}\partial_3\mathbf{k}^\varepsilon(\mathbf{x}-\gamma^0(r))\times\alpha'(r)\hat{\mathbf{x}}_2dr+\Gamma\int_{\mathbb{T}}\partial_3\mathbf{k}^\varepsilon(\mathbf{x}-\gamma^0(r))\times\hat{\mathbf{x}}_3dr.
\end{align*}
The first term is estimated (up to a constant) by $\Gamma\eta\varepsilon^{-2}$ using \eqref{e16a} and $|\alpha'(r)|\le\eta$. Finally, the same can be applied to the second one after observing that
\begin{align*}
\Gamma\int_{\mathbb{T}}\partial_3\mathbf{k}^\varepsilon(\mathbf{x}-\gamma^0(r))\times\hat{\mathbf{x}}_3dr&=-\Gamma\int_{\mathbb{T}}\frac{d}{dr}\mathbf{k}^\varepsilon(\mathbf{x}-\gamma^0(r))\times\hat{\mathbf{x}}_3dr-\Gamma\int_{\mathbb{T}}\partial_2\mathbf{k}^\varepsilon(\mathbf{x}-\gamma^0(r))\alpha'(r)\times\hat{\mathbf{x}}_3dr\\
&=-\Gamma\int_{\mathbb{T}}\partial_2\mathbf{k}^\varepsilon(\mathbf{x}-\gamma^0(r))\alpha'(r)\times\hat{\mathbf{x}}_3dr.
\end{align*}
\qed

We are finally able to show the theorem.

\medskip

{\sc Proof of Theorem~\ref{T:2}}
Taking the time derivative of \eqref{e1}, we obtain
\begin{align*}
\partial_t^2\gamma^{t=0}_2(r_0)=(\nabla\bar u_2)^{t=0}(\gamma^0(r_0))\cdot\partial_t\gamma^{t=0}(r_0)+\partial_t(v_2^t(\gamma^t(r_0))\big|_{t=0}+(\partial_t\bar u_2)^{t=0}(\gamma^0(r_0))
\end{align*}
The first term vanishes since $\bar u^0_2\equiv 0$. The second one satisfies the estimate
\begin{align}\label{e10}
|\partial_t(v_2^t(\gamma^t(r_0))\big|_{t=0}|\lesssim\eta\Big(\frac{\Gamma^2}{\varepsilon^3}+\frac{\Gamma\|\bar\omega^0\|_\infty}{\varepsilon}\Big),
\end{align}
which will be proven in the sequel. We can then discard it due to \eqref{e43}.

The main contribution will come from the last term, for which we have that
\begin{align*}
(\partial_t\bar u_2)^{t=0}(\gamma^0(r_0))&\overset{\eqref{e45}}{=}-\int_{\mathbb{T}^3}(\mathbf{k}^\varepsilon(\gamma^0(r_0)-\mathbf{y})\times\hat{\mathbf{x}}_2)\cdot(\partial_t\bar\omega)^{t=0}(\mathbf{y})dy\\
&\overset{\eqref{e44}}{=}\int_{\mathbb{T}^3}(\mathbf{k}^\varepsilon(\gamma^0(r_0)-\mathbf{y})\times\hat{\mathbf{x}}_2)\cdot (\bar{\mathbf{u}}^0\cdot\nabla\bar\omega^0-\bar\omega^0\cdot\nabla\bar{\mathbf{u}}^0)dy\\&+\int_{\mathbb{T}^3}(\mathbf{k}^\varepsilon(\gamma^0(r_0)-\mathbf{y})\times\hat{\mathbf{x}}_2)\cdot 
(\mathbf{v}^0\cdot\nabla\bar\omega^0-\bar\omega^0\cdot\nabla\mathbf{v}^0)dy\\
&\overset{\eqref{e3}}{=}\int_{\mathbb{T}^3}(\mathbf{k}^\varepsilon(\gamma^0(r_0)-\mathbf{y})\times\hat{\mathbf{x}}_2)\cdot 
(\mathbf{v}^0\cdot\nabla\bar\omega^0-\bar\omega^0\cdot\nabla\mathbf{v}^0)dy.
\end{align*}
We can neglect the second term (which takes into account the stretching effects) because
\begin{align*}
&\Big|\int_{\mathbb{T}^3}(\mathbf{k}^\varepsilon(\gamma^0(r_0)-\mathbf{y})\times\hat{\mathbf{x}}_2)\cdot 
(\bar\omega^0\cdot\nabla\mathbf{v}^0)dy\Big|\\&\overset{\eqref{e3}}{=}\Big|\int_{\mathbb{T}^3}(\mathbf{k}^\varepsilon(\gamma^0(r_0)-\mathbf{y})\times\hat{\mathbf{x}}_2)\cdot 
(\bar\omega^0_3\partial_3\mathbf{v}^0)dy\Big|\overset{\eqref{e18b}}{\lesssim}\frac{\Gamma\eta\|\bar\omega^0\|_\infty}{\varepsilon^2},
\end{align*}
where we also used that $\mathbf{k}^\varepsilon$ have uniformly bounded $L^1$ norm. Regarding the first term, using \eqref{e3}, it can be rewritten as
\begin{align*}
\partial_2\bar\omega^0_3(\gamma^0(r_0))\int_{\mathbb{T}^3}k_1^\varepsilon(\gamma^0(r_0)-\mathbf{y})v_2^0(\mathbf{y})dy+\int_{\mathbb{T}^3}k_1^\varepsilon(\gamma^0(r_0)-\mathbf{y})v^0_2(\mathbf{y})\big(\partial_2\bar\omega^0_3(\mathbf{y})-\partial_2\bar{\omega}^0_3(\gamma^0(r_0))\big)dy.
\end{align*}
The second term is an error, which satisfies the estimate
\begin{align}\label{e19}
\Big|\int_{\mathbb{T}^3}k_1^\varepsilon(\gamma^0(r_0)-\mathbf{y})v^0_2(\mathbf{y})\big(\partial_2\bar\omega^0_3(\mathbf{y})-\partial_2\bar{\omega}^0_3(\gamma^0(r_0))\big)dy\Big|\lesssim\Gamma\|\partial_2^2\bar\omega^0_3\|_\infty
\end{align}
proven below. The thesis will then follow from
\begin{align}\label{e21}
\int_{\mathbb{T}^3}k_1^\varepsilon(\gamma^0(r_0)-\mathbf{y})v_2^0(\mathbf{y})dy\sim\frac{\Gamma}{4\pi}\ln\frac{1}{\varepsilon}\qquad\mbox{as}~\varepsilon\to0.
\end{align}

We have
\begin{align}
\int_{\mathbb{T}^3}k_1^\varepsilon(\gamma^0(r)-\mathbf{y})v_2^0(\mathbf{y})dy&\overset{\eqref{e7},\eqref{e6}}{=}-\Gamma\int_{\mathbb{T}^3}\int_{\mathbb{T}}
k_1^\varepsilon(\gamma^0(r_0)-\mathbf{y})k_1^\varepsilon(\mathbf{y}-\gamma^0(r))drdy\nonumber\\
&=-\Gamma\int_{\mathbb{T}}(k_1^\varepsilon\ast k_1^\varepsilon)(\gamma^0(r_0)-\gamma^0(r))dr.\label{e9}
\end{align}
Using Lemma~\ref{L:2}, we learn that the main contribution to the last integral comes from the set $|r-r_0|\gg\varepsilon$. Indeed for $M=\ln^{\frac{1}{2}}(\varepsilon^{-1})$, we can split the last integral as
\begin{align*}
-\Gamma\int_{|r-r_0|\le M\varepsilon}k_1^\varepsilon\ast k_1^\varepsilon(\gamma^0(r_0)-\gamma^0(r))dr-\Gamma\int_{|r-r_0|> M\varepsilon}k_1^\varepsilon\ast k_1^\varepsilon(\gamma^0(r_0)-\gamma^0(r))dr.
\end{align*}
From Lemma~\ref{L:2}, we have $\|k_1^\varepsilon\ast k_1^\varepsilon\|_\infty\lesssim\varepsilon^{-1}$, which implies that the first term is $\Gamma\lesssim\ln^\frac{1}{2}(\varepsilon^{-1})$, hence is of lower order. Regarding the second term, it is equal to
\begin{align*}
&\frac{\Gamma}{4\pi}\int_{|r-r_0|>M\varepsilon}\frac{1}{|\gamma^0(r_0)-\gamma^0(r)|}\phi\Big(\frac{|\gamma^0(r_0)-\gamma^0(r)|}{\varepsilon}\Big)dr+O(\Gamma)\\
&=\Gamma\frac{1+o(1)}{4\pi}\int_{|r-r_0|>M\varepsilon}\frac{1}{|r-r_0|}dr+O(\Gamma)\qquad\mbox{(since $|\gamma^0(r_0)-\gamma^0(r)|\overset{\eta\ll1}{\sim}|r-r_0|\gg\varepsilon$)}\\
&=\Gamma\frac{1+o(1)}{4\pi}\ln\frac{1}{M\varepsilon}+O(\Gamma)=\Gamma\frac{1+o(1)}{4\pi}\ln\frac{1}{\varepsilon}+O(\Gamma).
\end{align*}
This concludes the proof of \eqref{e21}. It remains to prove the error estimates \eqref{e10} and \eqref{e19}.

Here comes the proof of \eqref{e10}. Starting from
\begin{align*}
v_2^t(\gamma^t(r_0))
&=-\Gamma\int_{\mathbb{T}}
(\mathbf{k}^\varepsilon(\gamma^t(r_0)-\gamma^t(r))\times\hat{\mathbf{x}}_2)\cdot
\partial_r\gamma^t(r)dr\\
&=-\Gamma\int_{\mathbb{T}} k^\varepsilon_1(\gamma^t(r_0)-\gamma^t(r))\partial_r\gamma^t_3(r)dr
+\Gamma\int_{\mathbb{T}} k^\varepsilon_3(\gamma^t(r_0)-\gamma^t(r))\partial_r\gamma^t_1(r)dr
\end{align*}
and taking the derivative, we obtain
\begin{align}\label{e13}
&\frac{d}{dt}v_2^t(\gamma^t(r_0))\Big|_{t=0}\nonumber\\
=&-\Gamma\int_{\mathbb{T}}\nabla
k_1^\varepsilon(\gamma^0(r_0)-\gamma^0(r))\cdot\big[\mathbf{v}^0(\gamma^0(r_0))-\mathbf{v}^0(\gamma^0(r))+
\bar{\mathbf{u}}^0(\gamma^0(r_0))-\bar{\mathbf{u}}^0(\gamma^0(r))\big]dr\nonumber\\
 &-\Gamma\int_{\mathbb{T}}
 k_1^\varepsilon(\gamma^0(r_0)-\gamma^0(r))\partial_r\big(v_3^0(\gamma^0(r))+\bar
 u_3^0(\gamma^0(r))\big)dr\nonumber\\
 &+\Gamma\int_{\mathbb{T}}
 k_3^\varepsilon(\gamma^0(r_0)-\gamma^0(r))\partial_r\big(v_1^0(\gamma^0(r))+\bar
 u_1^0(\gamma^0(r))\big)dr,
\end{align}
where we used that $\partial_r\gamma^0_3(r)=1$ and $\partial_r\gamma^0_1(r)=0$ (see~\eqref{e7}).

Let us notice the following cancellation. Using that
\begin{align*}
\partial_ik^\varepsilon_j(\mathbf{x})=(|\mathbf{x}|^2+\epsilon^2)^{-\frac{3}{2}}\delta_{ij}-3x_ix_j(|\mathbf{x}|^2+\epsilon^2)^{-\frac{5}{2}},\quad
\mbox{together with}~(\gamma^0(r_0)-\gamma^0(r))\cdot\hat{\mathbf{x}}_1\overset{\eqref{e7}}{=}0,
\end{align*}
we obtain
\begin{align}\label{e12}
\partial_2k_1^\varepsilon(\gamma^0(r_0)-\gamma^0(r))=\partial_3k^\varepsilon_3(\gamma^0(r_0)-\gamma^0(r))=0.
\end{align}

Starting from \eqref{e13}, let us integrate by parts the last two terms. Using \eqref{e12} and the fact that $\bar u^0_2=\bar u^0_3=0$ (cf.~\eqref{e3}), one derives the following
\begin{align*}
\frac{d}{dt}v_2(t,\gamma^t(r_0))\Big|_{t=0}
&=\Gamma\int_{\mathbb{T}}\partial_1k_1^\varepsilon(\gamma^0(r_0)-\gamma^0(r))[v_1^0(\gamma^0(r_0))-v_1^0(\gamma^0(r))]dr\\
&+\Gamma\int_{\mathbb{T}}(\alpha'(r)\partial_2k_3^\varepsilon+\partial_3k_3^\varepsilon)(\gamma^0(r_0)-\gamma^0(r))v_1^0(\gamma^0(r))dr\\
&+\Gamma\int_{\mathbb{T}}(\partial_1k_1^\varepsilon+\alpha'(r)\partial_2k_3^\varepsilon+\partial_3k_3^\varepsilon)\big(\gamma^0(r_0)-\gamma^0(r)\big)\big(\bar u_1^0(\gamma^0(r))-\bar
u_1^0(\gamma^0(r_0))\big)dr\\
&+\bar{u}_1^0(\gamma^0(r_0))\Gamma\int_{\mathbb{T}}\big(\partial_3k_3^\varepsilon(\gamma^0(r_0)-\gamma^0(r))+\alpha'(r)\partial_2k_3^\varepsilon(\gamma^0(r_0)-\gamma^0(r)\big)dr.
\end{align*}
Using \eqref{e16a} and \eqref{e18a}, the first two terms are $\lesssim\Gamma^2\eta\varepsilon^{-3}$. From
\begin{align*}
\big|\bar u_1^0(\gamma^0(r_0))-\bar u_1^0(\gamma^0(r))\big|\le\|(\bar u^0_1)'\|_\infty|\gamma^0_2(r_0)-\gamma^0_2(r)|\overset{\eqref{e41}}{\le}\eta\|(\bar u^0_1)'\|_\infty|r-r_0|\overset{\eqref{e42}}{=}\eta\|\bar\omega^0\|_\infty|r-r_0|
\end{align*}
and \eqref{e16b} from Lemma~\ref{L:1}, the third $\Gamma\eta\|\bar \omega^0\|_\infty\varepsilon^{-1}$. Finally, the last term is zero, since
\begin{align*}
\int_{\mathbb{T}}\big(\partial_3k_i^\varepsilon(\gamma^0(r_0)-\gamma^0(r))+\alpha'(r)\partial_2k_i^\varepsilon(\gamma^0(r_0)-\gamma^0(r)\big)dr=-\int_{\mathbb{T}}\partial_r\big(k_i^\varepsilon(\gamma^0(r_0)-\gamma^0(r))\big)dr=0.
\end{align*}

To conclude, let us establish \eqref{e19}. We claim that
\begin{align}\label{e20}
\int_{\mathbb{T}^3}\frac{1}{|\mathbf{x}-\mathbf{y}||\mathbf{y}|^2}dy\lesssim1+|\ln\frac{1}{|\mathbf{x}|}|
\end{align}
which follows from
\begin{align*}
\int_{\mathbb{T}^3}\frac{1}{|\mathbf{x}-\mathbf{y}||\mathbf{y}|^2}dy&\le\int_{B(0,1)}\frac{1}{|\mathbf{x}-\mathbf{y}||\mathbf{y}|^2}dy\overset{(\mathbf{y}=|\mathbf{x}|\mathbf{y}')}{=}\int_{B(0,\frac{1}{|\mathbf{x}|})}\frac{1}{|\frac{\mathbf{x}}{|\mathbf{\mathbf{x}}|}-\mathbf{y}'||\mathbf{y'}|^2}dy'\\&\lesssim1+\int_{B(0,\frac{1}{|\mathbf{x}|})\backslash B(0,2)}\frac{1}{|\mathbf{y}'|^3}dy'\lesssim1+\int_2^{\frac{1}{|\mathbf{x}|}}\frac{1}{\rho} d\rho\lesssim1+|\ln\frac{1}{|\mathbf{x}|}|.
\end{align*}
We then deduce \eqref{e19}
\begin{align*}
&\Big|\int_{\mathbb{T}^3}k_1^\varepsilon(\gamma^0(r_0)-\mathbf{y})v^0_2(\mathbf{y})\big(\partial_2\bar\omega^0_3(\mathbf{y})-\partial_2\bar{\omega}^0_3(\gamma^0(r_0))\big)dy\Big|\\
&\le\|\partial_2^2\bar\omega^0_3\|_\infty\int_{\mathbb{T}^3}|k_1^\varepsilon(\gamma^0(r_0)-\mathbf{y})||\gamma^0(r_0)-\mathbf{y}||v^0_2(\mathbf{y})|dy\\
&\overset{\eqref{e6}}{\lesssim}\Gamma\|\partial_2^2\bar\omega^0_3\|_\infty\int_{\mathbb{T}}\int_{\mathbb{T}^3}\frac{1}{|\gamma^0(r_0)-\mathbf{y}||\gamma^0(r)-\mathbf{y}|^2}dydr\\
&=\Gamma\|\partial_2^2\bar\omega^0_3\|_\infty\int_{\mathbb{T}}\int_{\mathbb{T}^3}\frac{1}{|\big(\gamma^0(r_0)-\gamma^0(r)\big)-\mathbf{y}||\mathbf{y}|^2}dydr\\&\overset{\eqref{e20}}{\lesssim}\Gamma\|\partial_2^2\bar\omega^0_3\|_\infty\int_{\mathbb{T}}\big(1+|\ln\frac{1}{|\gamma^0(r_0)-\gamma^0(r)|}|\big)dr\overset{\eqref{e7}}{\lesssim}\Gamma\|\partial_2^2\bar\omega^0_3\|_\infty.
\end{align*}
\qed

\subsection{The case of a thin domain}\label{S:thin}

Consider the set%
\[
D=\left[  0,1\right]  \times\left[  0,1\right]  \times\left[  0,\delta\right]
\]
and call by $\mathbb{T}_{\delta}^{3}$ the set obtained from $D$ by
identification of the boundaries as in the case of a torus;\ all functions on
$\mathbb{T}_{\delta}^{3}$ will be periodic, of period $1$ in the two
horizontal directions, period $\delta$ in the vertical direction.

In $\mathbb{T}_{\delta}^{3}$, consider a smooth curve of the form
\[
\gamma\left(  r\right)  =\alpha\left(  r\right)  \widehat{\mathbf{x}}%
_{2}+r\widehat{\mathbf{x}}_{3}%
\]
and a smooth background velocity field and the associated vorticity field,
given by
\[
\bar{\mathbf{u}}^{0}\left(  \mathbf{x}\right)  =\left(  \bar{u}%
_{1}^{0}\left(  x_{2}\right)  ,0,0\right)  ,\qquad\bar{\mathbf{\omega}%
}^{0}\left(  \mathbf{x}\right)  =\left(  0,0,\bar{\omega}_{3}^{0}\left(
x_{2}\right)  \right)  ,\qquad\mathbf{x}\in\mathbb{T}_{\delta}^{3}\mathbf{.}%
\]
Then introduce, on the classical torus $\mathbb{T}^{3}$, the curve%
\begin{align*}
\tilde{\gamma}\left(  s\right)    & =\tilde{\alpha}\left(  s\right)
\widehat{\mathbf{x}}_{2}+s\widehat{\mathbf{x}}_{3}\\
\tilde{\alpha}\left(  s\right)    & =\alpha\left(  \delta s\right)
\end{align*}
and the fields%
\[
\bar{\mathbf{u}}^{0,\delta}\left(  \mathbf{x}\right)  =\left(
\bar{u}_{1}^{0}\left(  x_{2}\right)  ,0,0\right)  ,\qquad\bar
{\mathbf{\omega}}^{0,\delta}\left(  \mathbf{x}\right)  =\left(  0,0,\bar
{\omega}_{3}^{0}\left(  x_{2}\right)  \right)  ,\qquad\mathbf{x}\in
\mathbb{T}^{3}.
\]
If the original fields depended on $x_{3}$, the new ones would have been
rescaled. But they were independent, hence we may drop the superscript
$\delta$ from $\bar{\mathbf{u}}^{0,\delta}$ and $\bar{\mathbf{\omega
}}^{0,\delta}$.

Given these initial conditions in the thin domain $\mathbb{T}_{\delta}^{3}$
and in the normalized domain $\mathbb{T}^{3}$, we may consider their time
evolution. For the sake of the following result we call $\left(  \gamma
^{t}\left(  r\right)  \right)  _{r}$ and $\left(  \tilde{\gamma}%
^{t}\left(  s\right)  \right)  _{s}$ the evolution at time $t$ of the curves
$\left(  \gamma\left(  r\right)  \right)  _{r}$ and $\left(  \tilde{\gamma
}\left(  s\right)  \right)  _{s}$ respectively; using the equations of motion
it is elementary to verify that $\partial_{t}^{2}\tilde{\gamma}_{2}%
^{t=0}\left(  s\right)  =\partial_{t}^{2}\gamma_{2}^{t=0}\left(  \delta
s\right)  $.

If we set%
\[
\eta:=\left\Vert \alpha^{\prime}\left(  r\right)  \right\Vert _{\infty}%
\]
then
\[
\left\Vert \tilde{\alpha}^{\prime}\left(  s\right)  \right\Vert _{\infty
}=\delta\eta.
\]
Theorem~\ref{T:2} applies to $\tilde{\gamma}$, $\bar{\mathbf{\omega}}^{0}$
and gives us the following result. Assume $\partial_{2}\bar{\omega}%
_{3}^{0}\left(  \tilde{\gamma}_{2}\left(  s_{0}\right)  \right)  >0$. In
the limit $\varepsilon\ll1$ and
\[
\delta\eta\ll\varepsilon^{3}\frac{\partial_{2}\bar{\omega}_{3}^{0}\left(
\tilde{\gamma}_{2}\left(  s_{0}\right)  \right)  }{\Gamma}%
\]
we have%
\[
\partial_{t}^{2}\tilde{\gamma}_{2}^{t=0}\left(  s_{0}\right)  \sim
\frac{\Gamma\partial_{2}\bar{\omega}_{3}^{0}\left(  \tilde{\gamma
}_{2}\left(  s_{0}\right)  \right)  }{4\pi}\ln\frac{1}{\varepsilon}.
\]
Call $r_{0}=\delta s_{0}$. Then we have the following result in the thin
domain $\mathbb{T}_{\delta}^{3}$:

\begin{cor}
If $\partial_{2}\bar{\omega}_{3}^{0}\left(  \gamma_{2}\left(
r_{0}\right)  \right)  >0$, in the limit $\varepsilon\ll1$ and
\[
\delta\eta\ll\varepsilon^{3}\frac{\partial_{2}\bar{\omega}_{3}^{0}\left(
\gamma_{2}\left(  r_{0}\right)  \right)  }{\Gamma}%
\]
we have%
\[
\partial_{t}^{2}\gamma_{2}^{t=0}\left(  r_{0}\right)  \sim\frac{\Gamma
\partial_{2}\bar{\omega}_{3}^{0}\left(  \gamma_{2}\left(  r_{0}\right)
\right)  }{4\pi}\ln\frac{1}{\varepsilon}.
\]
\end{cor}

\begin{rmk}
In the standard torus, we need (up to constants) $\eta\ll\varepsilon^{3}$, which
looks quite severe, but we have to notice that we deal with a modification of a
vertical straight line (namely a modification of the pure 2D case) which is
global, affecting the whole length of the domain. In the case of a $\delta
$-thin domain, which is common in geophysical fluid dynamics and a very natural set-up for a quasi-2D theory, the
condition becomes $\delta\eta\ll\varepsilon^{3}$ (up to constants), which is less severe.
\end{rmk}

\section{Conclusions}

With reference to the conceptual subdivision in the classes (a) and (b) of
vortex structure interactions described in the Introduction, we focused on
case (b), precisely on a concentrated 2D vortex structure of small radius in
background shear flow, with a non-zero gradient of vorticity. We have proved a
rigorous theorem about the motion along the gradient, confirming without any
approximation a phenomenon previously identified on experimental and numerical
basis, with certain approximate theoretical justifications. We have reached
the explicit formula for the displacement in the orthogonal direction to the
fluid flow
\begin{equation}
G_{2}\left(  t\right)  \sim\left(  \frac{\Gamma\partial_{2}\bar{\omega
}^{0}\left(  \mathbf{x}^{0}\right)  }{4\pi}\ln\frac{1}{\varepsilon}\right)
t^{2}\label{rigorous in summary}%
\end{equation}
The first derivative is zero and the second derivative diverges as the radius
$\varepsilon$ of the vortex structure goes to zero (vortex point limit);
numerically, this is confirmed by the log-log plots of the graph of
$G_{2}\left(  t\right)  $, which in the limit as $\varepsilon\rightarrow0$
behaves qualitatively as $G_{2}\left(  t\right)  \sim t^{\alpha}$ with powers
$\alpha$ less than 2. A very vague fit gives $\alpha=\frac{3}{2}$ in an
intermediate stage, before going down towards $\alpha=1$ (or even less, but
when time is so large that the peculiarities of the deformation of the
background do not allow any more to deduce general principles). In spite of
many attempts, we have not found an explanation of the power $\alpha=\frac
{3}{2}$;\ therefore we are not sure if it is a true result or only the average
indication of values smaller than 2, slowly changing around $\frac{3}{2}$.

Concerning the quasi-2D-flows, we have proved that a quasi-straight-line
vortex filament in a shear flow performs a motion along the gradient, as in
the 2D case. In order to handle such complex 3D system we have idealized a
realistic vortex tube as a vortex filament (a curve) and we have paid the
price of introducing a mollification of the Biot-Savart kernel at scale
$\varepsilon$, the number that should describe the average radius of the tube.
Three main open problems arose that we cannot treat in this work: a)
generalize to a realistic vortex tube; b) find a threshold, like the one
discussed in \cite{Alexakis}, beyond which the behaviour is different; c)
identify special features, typical of the 3D case, even in the case under
consideration where the motion along the gradient holds. In particular, about
(c), from heuristic consideration we have the following conjecture.

i) where $\alpha^{\prime\prime}\left(  r\right)  <0$, a larger concentration
of background vorticity is created \textquotedblright behind\textquotedblright%
\ the vortex filament, which produces a stronger displacement, locally, of the
filament along the gradient; with opposite behaviour where $\alpha
^{\prime\prime}\left(  r\right)  >0$;

ii) with the consequence of the beginning of an instability of the filament,
eventually leading to an increase of deformation with respect to a vertical
straight line.

However, the self-interaction of the vortex filament is so strong that perhaps
the previous phenomena, even if existing, could be less relevant, and
instabilities could occur more strongly for other reasons.

Among the perspective of our work there a rigorous analysis of stages of the
dynamics which go beyond the first one, and analytical computations beyond the
assumption of a small vorticity blob (clump or hole) in a background or larger
vortex, approaching the case of vortices approximately of the same size.
However it seems that these two problems require new techniques.

\section*{Acknowledgement}
The research of F.F. is funded by the European Union, ERC NoisyFluid, No. 101053472 and the research project PRIN 2022 “Noise in fluid dynamics and related models”, no. 20222YRYSP. We thank the anonymous referees for an exceptionally deep list of remarks which was the basis for a massive reformulation of the manuscript and the addition of many results, in particular the quasi-two-dimensional case.

\bibliographystyle{plain}
\bibliography{biblio.bib}

\begin{thebibliography}{10}

\bibitem{Alexakis}
Alexandros Alexakis.
\newblock Quasi-two-dimensional turbulence.
\newblock {\em Reviews of Modern Plasma Physics}, 7(1):31, 2023.

\bibitem{Amoretti}
M~Amoretti, D~Durkin, J~Fajans, R~Pozzoli, and M~Rom{\'e}.
\newblock Asymmetric vortex merger: Experiments and simulations.
\newblock {\em Physics of Plasmas}, 8(9):3865--3868, 2001.

\bibitem{Banica2}
Valeria Banica, Daniel Eceizabarrena, Andrea~R Nahmod, and Luis Vega.
\newblock Multifractality and intermittency in the limit evolution of polygonal
  vortex filaments.
\newblock {\em Mathematische Annalen}, 391(2):2837--2899, 2025.

\bibitem{Banica1}
Valeria Banica and Luis Vega.
\newblock Evolution of polygonal lines by the binormal flow.
\newblock {\em Annals of PDE}, 6(1):6, 2020.

\bibitem{Benedetto}
Dario Benedetto, Emanuele Caglioti, and Carlo Marchioro.
\newblock On the motion of a vortex ring with a sharply concentrated vorticity.
\newblock {\em Mathematical methods in the applied sciences}, 23(2):147--168,
  2000.

\bibitem{berselli}
Luigi~C Berselli and Hakima Bessaih.
\newblock Some results for the line vortex equation.
\newblock {\em Nonlinearity}, 15(6):1729, 2002.

\bibitem{Bessaih}
Hakima Bessaih, Michele Coghi, and Franco Flandoli.
\newblock Mean field limit of interacting filaments and vector valued
  non-linear pdes.
\newblock {\em Journal of Statistical Physics}, 166(5):1276--1309, 2017.

\bibitem{Boffetta}
Guido Boffetta and Robert~E Ecke.
\newblock Two-dimensional turbulence.
\newblock {\em Annual review of fluid mechanics}, 44(1):427--451, 2012.

\bibitem{Bog1977}
V.~Bogomolov.
\newblock Dynamics of vorticity at a sphere.
\newblock {\em Fluid Dyn.}, 6:863--870, 1977.

\bibitem{Briggs}
RJ~Briggs, JD~Daugherty, and RH~Levy.
\newblock Role of landau damping in crossed-field electron beams and inviscid
  shear flow.
\newblock {\em The Physics of Fluids}, 13(2):421--432, 1970.

\bibitem{Butta}
Paolo Butt{\`a} and Carlo Marchioro.
\newblock Long time evolution of concentrated euler flows with planar symmetry.
\newblock {\em SIAM Journal on Mathematical Analysis}, 50(1):735--760, 2018.

\bibitem{Caglioti}
Emanuele Caglioti, Pierre-Louis Lions, Carlo Marchioro, and Mario Pulvirenti.
\newblock A special class of stationary flows for two-dimensional euler
  equations: a statistical mechanics description. part ii.
\newblock {\em Communications in Mathematical Physics}, 174:229--260, 1995.

\bibitem{Carnevale}
George~F. Carnevale, Rudolf~C. Kloosterziel, and Gertjan J.~F. Van~Heijst.
\newblock Propagation of barotropic vortices over topography in a rotating
  tank.
\newblock {\em Journal of Fluid Mechanics}, 233:119--139, 1991.

\bibitem{Chan}
Johnny~CL Chan and RT~Williams.
\newblock Analytical and numerical studies of the beta-effect in tropical
  cyclone motion. part i: Zero mean flow.
\newblock {\em Journal of atmospheric sciences}, 44(9):1257--1265, 1987.

\bibitem{Chavanis}
Pierre-Henri Chavanis.
\newblock Statistical mechanics of two-dimensional point vortices: relaxation
  equations and strong mixing limit.
\newblock {\em The European Physical Journal B}, 87:1--24, 2014.

\bibitem{Fefferman}
Diego Cordoba and Charles Fefferman.
\newblock On the collapse of tubes carried by 3d incompressible flows.
\newblock {\em arXiv preprint math/0101253}, 2001.

\bibitem{Darios}
LS~Da~Rios.
\newblock On the motion of an unbounded fluid with a vortex filament of any
  shape.
\newblock {\em Rend}, 22:117, 1906.

\bibitem{DelaHoz}
Francisco De~la Hoz and Luis Vega.
\newblock Vortex filament equation for a regular polygon.
\newblock {\em arXiv preprint arXiv:1304.5521}, 2013.

\bibitem{Diamond}
Patrick~H. Diamond, S.~I. Itoh, K.~Itoh, and T.~S. Hahm.
\newblock Zonal flows in plasma—a review.
\newblock {\em Plasma Physics and Controlled Fusion}, 47(5):R35, 2005.

\bibitem{Dritschel}
David~G Dritschel.
\newblock A general theory for two-dimensional vortex interactions.
\newblock {\em Journal of Fluid Mechanics}, 293:269--303, 1995.

\bibitem{Dupree}
Thomas~H Dupree.
\newblock Growth of phase-space density holes.
\newblock {\em The Physics of fluids}, 26(9):2460--2481, 1983.

\bibitem{Eyink}
GL~Eyink and H~Spohn.
\newblock Negative-temperature states and large-scale, long-lived vortices in
  two-dimensional turbulence.
\newblock {\em Journal of statistical physics}, 70:833--886, 1993.

\bibitem{Eyink2}
Gregory~L Eyink and Katepalli~R Sreenivasan.
\newblock Onsager and the theory of hydrodynamic turbulence.
\newblock {\em Reviews of modern physics}, 78(1):87--135, 2006.

\bibitem{Lau2005}
Laurent-Polz F.
\newblock Point vortices on a rotating sphere.
\newblock {\em Regular and Chaotic Dynamics}, 10(1):39--58, 2005.

\bibitem{Flandoli}
Franco Flandoli.
\newblock Renormalized onsager functions and merging of vortex clusters.
\newblock {\em Stochastics and Dynamics}, 20(06):2040010, 2020.

\bibitem{Fontelos}
Marco~A Fontelos and Luis Vega.
\newblock Evolution of viscous vortex filaments and desingularization of the
  biot-savart integral.
\newblock {\em arXiv preprint arXiv:2311.12246}, 2023.

\bibitem{Fukumoto}
Yasuhide Fukumoto and Takeshi Miyazaki.
\newblock Three-dimensional distortions of a vortex filament with axial
  velocity.
\newblock {\em Journal of fluid mechanics}, 222:369--416, 1991.

\bibitem{Garbet}
Xavier Garbet, Yasuhiro Idomura, Laurent Villard, and TH~Watanabe.
\newblock Gyrokinetic simulations of turbulent transport.
\newblock {\em Nuclear Fusion}, 50(4):043002, 2010.

\bibitem{Hasegawa}
Akira Hasegawa and Kunioki Mima.
\newblock Stationary spectrum of strong turbulence in magnetized nonuniform
  plasma.
\newblock {\em Physical Review Letters}, 39(4):205, 1977.

\bibitem{Horton}
Wendell Horton and Akira Hasegawa.
\newblock Quasi-two-dimensional dynamics of plasmas and fluids.
\newblock {\em Chaos: An Interdisciplinary Journal of Nonlinear Science},
  4(2):227--251, 1994.

\bibitem{Jerrard}
Robert~L Jerrard and Christian Seis.
\newblock On the vortex filament conjecture for euler flows.
\newblock {\em Archive for Rational Mechanics and Analysis}, 224(1):135--172,
  2017.

\bibitem{Keener}
James~P Keener.
\newblock Knotted vortex filaments in an ideal fluid.
\newblock {\em Journal of Fluid Mechanics}, 211:629--651, 1990.

\bibitem{Krajchnan}
Fobert~H Krajchnan.
\newblock Inertial ranges in two-dimensional turbulence.
\newblock {\em The Physics of Fluids}, 10:1417, 1967.

\bibitem{Lansky}
IM~Lansky, TM~O'Neil, and DA~Schecter.
\newblock A theory of vortex merger.
\newblock {\em Physical review letters}, 79(8):1479, 1997.

\bibitem{LeviCivita}
Tullio Levi-Civita.
\newblock Teoremi di unicita e di esistenza per le piccole oscillazioni di un
  filetto vorticoso prossimo alla forma circolare (uniqueness and existence
  theorems for small oscillations of a vortex filament near the circular shape.
\newblock {\em Rend. R. Acc. Lincei}, 15:409--416, 1932.

\bibitem{Li}
Xiaofan Li and Bin Wang.
\newblock Barotropic dynamics of the beta gyres and beta drift.
\newblock {\em Journal of Atmospheric Sciences}, 51(5):746--756, 1994.

\bibitem{Lions}
Pierre-Louis Lions.
\newblock {\em On Euler equations and Statistical Physics}.
\newblock Scuola Normale Superiore (Edizioni della Normale, 1997).

\bibitem{Majda}
Andrew~J. Majda and Andrea~L. Bertozzi.
\newblock {\em Vorticity and Incompressible Flow}.
\newblock Cambridge Texts in Applied Mathematics. Cambridge University Press,
  2001.

\bibitem{MarcPulv}
Carlo Marchioro and Mario Pulvirenti.
\newblock On the vortex--wave system.
\newblock In {\em Mechanics, analysis and geometry: 200 years after Lagrange},
  pages 79--95. Elsevier, 1991.

\bibitem{Marchioro}
Carlo Marchioro and Mario Pulvirenti.
\newblock {\em Mathematical theory of incompressible nonviscous fluids},
  volume~96.
\newblock Springer Science \& Business Media, 2012.

\bibitem{Marcus2004}
Philip~S Marcus.
\newblock Prediction of a global climate change on jupiter.
\newblock {\em Nature}, 428(6985):828--831, 2004.

\bibitem{Marcus2000}
Philip~S Marcus, T~Kundu, and Changhoon Lee.
\newblock Vortex dynamics and zonal flows.
\newblock {\em Physics of Plasmas}, 7(5):1630--1640, 2000.

\bibitem{Mariotti}
Annarita Mariotti, Bernard Legras, and David~G Dritschel.
\newblock Vortex stripping and the erosion of coherent structures in
  two-dimensional flows.
\newblock {\em Physics of Fluids}, 6(12):3954--3962, 1994.

\bibitem{ModViv2020}
Klas Modin and Milo Viviani.
\newblock A casimir preserving scheme for long-time simulation of spherical
  ideal hydrodynamics.
\newblock {\em Journal of Fluid Mechanics}, 884(22A), 2020.

\bibitem{ModViv2025}
Klas Modin and Milo Viviani.
\newblock A brief introduction to matrix hydrodynamics.
\newblock {\em arXiv preprint arXiv:2508.07088}, 2025.

\bibitem{Montgomery}
David Montgomery and Glenn Joyce.
\newblock Statistical mechanics of negative temperature states.
\newblock 1973.

\bibitem{MooreSaffman}
DW~Moore and P~Gr Saffman.
\newblock Structure of a line vortex in an imposed strain.
\newblock pages 339--354, 1971.

\bibitem{Nielsen}
AH~Nielsen, X~He, J~Juul Rasmussen, and T~Bohr.
\newblock Vortex merging and spectral cascade in two-dimensional flows.
\newblock {\em Physics of Fluids}, 8(9):2263--2265, 1996.

\bibitem{Ogawa}
Akihito Ogawa, Akio Sanpei, and Haruhiko Himura.
\newblock Convective vortex merging in electron plasma.
\newblock {\em Physics of Plasmas}, 32(6), 2025.

\bibitem{Onsager}
Lars Onsager.
\newblock Statistical hydrodynamics.
\newblock {\em Il Nuovo Cimento (1943-1954)}, 6(Suppl 2):279--287, 1949.

\bibitem{Ozugurlu}
Ersin {\"O}zu{\u{g}}urlu, Jean~N Reinaud, and David~G Dritschel.
\newblock Interaction between two quasi-geostrophic vortices of unequal
  potential vorticity.
\newblock {\em Journal of Fluid Mechanics}, 597:395--414, 2008.

\bibitem{Reinaud}
Jean~Noel Reinaud and David~Gerard Dritschel.
\newblock The merger of vertically offset quasi-geostrophic vortices.
\newblock {\em Journal of Fluid Mechanics}, 469:287--315, 2002.

\bibitem{Rhines}
Peter~B Rhines.
\newblock Waves and turbulence on a beta-plane.
\newblock {\em Journal of Fluid Mechanics}, 69(3):417--443, 1975.

\bibitem{Ricca}
Renzo~L Ricca.
\newblock Rediscovery of da rios equations.
\newblock {\em Nature}, 352(6336):561--562, 1991.

\bibitem{Saffman}
Philip~G Saffman.
\newblock Vortex dynamics.
\newblock In {\em Theoretical Approaches to Turbulence}, pages 263--277.
  Springer, 1992.

\bibitem{Sanchez}
A~S{\'a}nchez-Lavega, JF~Rojas, R~Hueso, J~Lecacheux, F~Colas, JR~Acarreta,
  I~Miyazaki, and D~Parker.
\newblock Interaction of jovian white ovals bc and de in 1998 from earth-based
  observations in the visual range.
\newblock {\em Icarus}, 142(1):116--124, 1999.

\bibitem{Schecter}
David~A Schecter and Daniel~HE Dubin.
\newblock Vortex motion driven by a background vorticity gradient.
\newblock {\em Physical review letters}, 83(11):2191, 1999.

\bibitem{Vallis}
Geoffrey~K Vallis.
\newblock {\em Atmospheric and oceanic fluid dynamics}.
\newblock Cambridge University Press, 2017.

\bibitem{Vasavada}
Ashwin~R Vasavada and Adam~P Showman.
\newblock Jovian atmospheric dynamics: An update after galileo and cassini.
\newblock {\em Reports on Progress in Physics}, 68(8):1935, 2005.

\bibitem{Hardenberg}
J~Von~Hardenberg, JC~McWilliams, A~Provenzale, A~Shchepetkin, and JB~Weiss.
\newblock Vortex merging in quasi-geostrophic flows.
\newblock {\em Journal of Fluid Mechanics}, 412:331--353, 2000.

\bibitem{Wong}
Michael~H Wong, Imke de~Pater, Xylar Asay-Davis, Philip~S Marcus, and
  Christopher~Y Go.
\newblock Vertical structure of jupiter’s oval ba before and after it
  reddened: What changed?
\newblock {\em Icarus}, 215(1):211--225, 2011.

\bibitem{Yudovich}
Victor~I Yudovich.
\newblock Non-stationary flow of an ideal incompressible liquid.
\newblock {\em USSR Computational Mathematics and Mathematical Physics},
  3(6):1407--1456, 1963.

\bibitem{Zei2004}
Vladimir Zeitlin.
\newblock Self-consistent finite-mode approximations for the hydrodynamics of
  an incompressible fluid on nonrotating and rotating spheres.
\newblock {\em Phys. Rev. Lett.}, 93(26):264--501, 2004.

\bibitem{Zhou2017}
Ye~Zhou.
\newblock Rayleigh--taylor and richtmyer--meshkov instability induced flow,
  turbulence, and mixing. ii.
\newblock {\em Physics Reports}, 723:1--160, 2017.

\bibitem{Zhou2024}
Ye~Zhou.
\newblock {\em Hydrodynamic Instabilities and Turbulence: Rayleigh--Taylor,
  Richtmyer--Meshkov, and Kelvin--Helmholtz Mixing}.
\newblock Cambridge University Press, 2024.

\bibitem{Zhou2025}
Ye~Zhou, James~D Sadler, and Omar~A Hurricane.
\newblock Instabilities and mixing in inertial confinement fusion.
\newblock {\em Annual Review of Fluid Mechanics}, 57:197--225, 2025.

\end{thebibliography}

\end{document}